\documentclass[twocolumn]{aa}
\usepackage{natbib}
\usepackage{color}
\usepackage{ragged2e}
\usepackage[pdfpagelabels=false]{hyperref}	
\hypersetup{colorlinks=true,linkcolor=blue,citecolor=blue,filecolor=blue,urlcolor=blue,}
\usepackage[varg]{txfonts}
\usepackage{graphicx,rotating}
\usepackage[normalem]{ulem}
\usepackage{colortbl}
\usepackage{xcolor}
\usepackage{bbold}
\usepackage{longtable}
\usepackage{booktabs}
\usepackage{array}
\usepackage{listings}
\usepackage{xcolor}

\lstdefinestyle{pythonstyle}{
    language=Python,
    basicstyle=\ttfamily\scriptsize,
    keywordstyle=\color{blue},
    commentstyle=\color{gray},
    stringstyle=\color{teal},
    numbers=none,
    breaklines=true,
    breakatwhitespace=false,
    showstringspaces=false,
    frame=none,
    tabsize=4,
    columns=fullflexible
}

\bibpunct{(}{)}{;}{a}{}{,} 

\definecolor{darkolivegreen}{rgb}{0.33, 0.42, 0.18}
\definecolor{salmon}{rgb}{0.95,0.5,0.25}
\newcommand{\pb}[1]{\textbf{ \textcolor{magenta}{Pierre: #1}}}

\begin{document}

\title{\texttt{CosmoDyn}: a semi-analytic post-processing framework for the dynamical evolution of compact stellar systems and black holes in cosmological simulations}

\titlerunning{CosmoDyn}
\author{Pierre Boldrini \inst{1}, Paola Di Matteo \inst{1}, David Katz \inst{1}, Laia Casamiquela \inst{1}, Chervin Laporte \inst{1,2,3}, Maxime Assuerus \inst{1}, Marco Montuori\inst{4,5} and Glenn van de Ven\inst{6}}

\offprints{Pierre Boldrini, \email{pierre.boldrini@obspm.fr}}
\institute{$^{1}$ LIRA, Observatoire de Paris, Université PSL, Sorbonne Université, Université Paris Cité, CY Cergy Paris Université, CNRS, 92190 Meudon, France, \\
$^{2}$ Kavli IPMU (WPI), UTIAS, The University of Tokyo, Kashiwa, Chiba 277-8583, Japan, \\
$^{3}$ Institut de Ciències del Cosmos (ICCUB), Universitat de Barcelona, Martí i Franquès 1, E-08028 Barcelona, Spain, \\
$^{4}$ Institute for Complex Systems - Consiglio Nazionale delle Ricerche, Piazzale A. Moro 2, I-00185, Rome, Italy, \\
$^{5}$ Physics Department, Sapienza Università di Roma, Piazzale A. Moro 2, I-00185, Rome, Italy \\
$^{6}$ Department of Astrophysics, University of Vienna, T\"urkenschanzstra{\ss}e 17, 1180 Vienna, Austria}
\authorrunning{Boldrini et al.}
\date{submitted to A$\&$A}

\abstract{

Although a large number of cosmological simulations are now publicly available, their spatial and mass resolutions remain insufficient to accurately follow the dynamics of compact stellar systems and black holes. A complementary approach is therefore required to exploit the cosmological assembly histories encoded in these simulations without rerunning them. We present \texttt{CosmoDyn}, an open-source Python framework that reconstructs time-dependent galactic environments from any cosmological simulations and explicitly follows unresolved compact objects within them. A distinctive feature of \texttt{CosmoDyn} is that it models not only the main host galaxy but also its accreted satellites via moving gravitational potentials. The framework can therefore account simultaneously for the delivery of ex-situ populations and for the dynamical influence of satellites on in-situ population already present in the host. The modular pipeline comprises four main stages: reconstructing the evolving host and satellite potentials; generating in-situ and ex-situ populations of compact objects; integrating their orbits with configurable prescriptions for dynamical friction, and mass loss; and modeling tidal streams. Once reconstructed, the same cosmological environment can be reused to explore different objects, initial conditions, and physical prescriptions at low computational cost. \texttt{CosmoDyn} thus enables rapid parameter exploration and applications to statistically significant galaxy samples, while its modular architecture is designed to incorporate progressively more realistic potentials, and formation models.

}

\maketitle




\section{Introduction}

The dynamical evolution of compact stellar systems and black holes (BHs) is entering a data-rich era. The \textit{Gaia} mission has transformed our view of the Milky Way (MW) by providing precise phase-space information for its stellar populations, globular clusters (GCs), and tidal streams \citep{Gaia21}, as well as unprecedented samples of open clusters \citep{Laia24}. In parallel, \textit{Euclid} is extending these studies to the extragalactic regime by detecting large populations of GCs and nuclear star clusters (NSCs) across a wide range of galaxies and environments. The Early Release Observations already provide several thousand GC and NSC candidates in the Fornax and Perseus galaxy clusters \citep{saifollahi2025_fornax,saifollahi2025_perseus}, while the number of robust GC candidates expected within the regions observed by \textit{Euclid} out to approximately 70~Mpc may approach $10^5$ \citep{Lancon21,Voggel25}. Beyond tracing the assembly of their host galaxies, NSCs are intimately connected to the formation and growth of BHs, which can coexist with them in galactic nuclei or be left wandering after the disruption of their host galaxies. The \textit{LISA} mission, planned for launch in 2035, will open a complementary window on massive and intermediate-mass BHs through gravitational-wave observations \citep{LISA}.

Studying the dynamics of these objects over cosmological timescales and  within statistically representative galaxy populations is essential for understanding the formation and assembly of galaxies. However, simultaneously resolving the scales of these objects and the cosmological environments in which they evolve remains computationally challenging \citep[for reviews, see][]{Beasley20,Renaud20}. Their characteristic masses and sizes are generally below the mass and spatial resolutions of large cosmological simulations, whereas simulations capable of resolving their internal structures are necessarily restricted in volume, duration, or galaxy sample size.

Several complementary strategies have been developed to overcome these limitations. Idealised galaxy simulations and high-resolution cosmological zoom-in simulations can reach spatial resolutions of approximately $1$--$10$~pc and directly study clustered star formation \citep{Kravtsov05,Li17,Kim18,Ma20,Meng22,Sameie23,Dubois21,Lahen20,Li22,Deng24,Andersson24}, but their computational cost restricts them to small galaxy samples, limited volumes, or high redshifts. Semi-analytical cluster-formation and evolution models incorporated into hydrodynamical simulations follow GCs in a realistic cosmological context \citep{Pfeffer18,ReinaCampos22,Grudic23,Newton24,Rodriguez23}, but remain limited by the cost of the underlying simulations. Post-processing tagging methods can instead populate thousands of galaxies without rerunning the simulations \citep{Bullock05,Cooper10,Laporte13,Renaud17,Ramos-Almendares20,Halbesma20,Park22,Doppel23,Chen23,Creasey19}. However, the dynamics of the tagged objects are generally inherited from stellar or dark matter (DM) particles whose masses, spatial resolutions, and dynamical responses may differ substantially from those of the compact systems they represent. Similar numerical limitations affect BHs. In many cosmological simulations, BH particles are artificially repositioned toward the local potential minimum to prevent numerical scattering and maintain efficient sub-grid accretion and feedback models \citep{Weinberger2017MNRAS.465.3291W,bahe2021_https://doi.org/10.48550/arxiv.2109.01489}. Although computationally convenient, this procedure suppresses physically meaningful off-centre BH populations, bypasses unresolved dynamical friction, and can overestimate BH growth and merger rates \citep{Barausse2020ApJ...904...16B,bahe2021_https://doi.org/10.48550/arxiv.2109.01489}. More physical prescriptions for unresolved dynamical friction have been introduced \citep[e.g.][]{tremmel2015MNRAS.451.1868T,2019MNRAS.486..101P,Chen2022MNRAS.510..531C}, but following off-centre and wandering BHs over cosmological timescales and across large galaxy samples remains challenging.

We introduce \texttt{CosmoDyn} to bridge the gap between the spatial and temporal resolution of cosmological simulations and the dynamical scales of compact objects. In post-processing, \texttt{CosmoDyn} reconstructs the time-dependent gravitational environment extracted from a cosmological simulation using analytical potentials, adds unresolved populations of GCs, open clusters, NSCs, or BHs, and explicitly integrates their orbits while accounting for the selected physical processes. This approach provides three main advantages. First, it allows objects whose masses and sizes are unresolved by the original cosmological simulation to be represented with their appropriate physical properties. Second, because all calculations are performed in post-processing, the properties and evolutionary prescriptions of the objects can be modified without rerunning the computationally expensive cosmological hydrodynamical simulation. Different gravitational potentials, dynamical-friction models, mass-loss prescriptions, and initial populations can therefore be explored rapidly within the same cosmological assembly history. Third, its low computational cost makes it possible to analyse statistically significant samples of simulated systems, as required to interpret the large present and forthcoming observational datasets. The scope of \texttt{CosmoDyn} is deliberately intermediate between direct numerical simulations and conventional particle-tagging methods. It is designed to follow tagged objects whose gravitational influence on the global evolution of their environment remains negligible, such that they can be treated as test particles, extended perturbers, or point masses moving within an externally imposed potential. The code does not model the hydrodynamical formation of these objects, their feedback on their environment, or the self-consistent response of the gravitational potential to their orbital evolution. The current implementation is intended for hierarchically structured environments that can be represented by a dominant host potential and a population of lower-mass satellites or substructures. The dominant host may correspond, for example, to an individual galaxy or to a galaxy cluster, allowing the dynamics of compact objects to be followed both within galaxies and within the satellite galaxies of a larger system. However, \texttt{CosmoDyn} is not currently designed to describe mergers between comparable-mass systems, during which the decomposition into a dominant host and lower-mass satellites may no longer be appropriate.

In this paper, we present the physical framework, numerical implementation, inputs, outputs, and main options of \texttt{CosmoDyn}. In Section~\ref{sec:full_potential}, we describe how \texttt{CosmoDyn} reconstructs the time-dependent gravitational potentials of the main host galaxy and its satellites, and illustrate this procedure using a MW-like galaxy selected from TNG50. Section~\ref{sec:initial_conditions} presents the generation of the in-situ and ex-situ GC, NSC, and BH populations. In Section~\ref{sec:dynamics}, we describe their orbital integration and the implementation of classical, anisotropic, and fuzzy dark matter (FDM) dynamical friction, mass loss, and tidal-stream modelling. The input and output files, a complete launcher reproducing our application, and a concise getting-started guide are provided in Appendices~\ref{ApA} and ~\ref{ApB}. The source code, example input files, and documentation are publicly available at \url{https://github.com/Blackholan/CosmoDyn}. This paper refers to code version 1.2.

\section{Cosmological environment}\label{sec:full_potential}

The main objective of the current version of \texttt{CosmoDyn} is to reconstruct, outside a cosmological simulation, its time-dependent galactic environments as a superposition of analytical gravitational potentials whose structural parameters evolve with time. Future versions will extend this framework to more realistic mass distributions using basis-function expansion methods. In a cosmological context, the gravitational potential is composed of a central host galaxy together with its accreted and surviving satellite galaxies throughout cosmic time. Our approach has been designed to be applicable to any cosmological simulation. Several publicly available hydrodynamical cosmological simulations can be used for this purpose, including TNG50\footnote{\url{https://www.tng-project.org/}}, EAGLE\footnote{\url{https://astro.dur.ac.uk/ICC/Eagle/database.php}}, FIRE-2 DR2\footnote{\url{https://flathub.flatironinstitute.org/fire}}, and Auriga\footnote{\url{https://wwwmpa.mpa-garching.mpg.de/auriga/data_new.html}}. This is a non-exhaustive list of publicly simulations suitable for studying MW-like galaxies with \texttt{CosmoDyn}. A key requirement is access either to the particle distribution shaping the gravitational potential, namely DM, stars, and gas at every simulation snapshot, or to the structural parameters of these components derived from halo and subhalo catalogues or computed directly from the simulation outputs. 

As an illustrative example, throughout this work we use the TNG50 cosmological simulation \citep{Nelson19b,Nelson19a,Pillepich19}, for which the structural properties of both DM haloes and stellar components are provided for every galaxy at each snapshot. These quantities enable a straightforward implementation of \texttt{CosmoDyn}. In other simulations, only the snapshots and halo catalogues identified with a halo finder may be available. In such cases, the merger trees linking galaxies across cosmic time must either be reconstructed by the user or obtained separately. Fortunately, merger trees are publicly available for all the listed simulations, considerably simplifying their exploitation. Throughout this paper, we will also discuss the differences that can be expected when applying \texttt{CosmoDyn} to simulations other than TNG50. In particular, the number of MW analogues, the achievable resolution, and the adopted simulation strategy (large cosmological volume versus zoom-in simulations) may all influence the resulting statistical analyses.

\section{\texttt{CosmoDyn} overview and workflow}

\texttt{CosmoDyn} is controlled through a Python launcher that specifies the galaxies to be analysed and the physical and numerical configuration of the calculation. It relies extensively on the publicly available \texttt{galpy} library\footnote{\url{https://github.com/jobovy/galpy}} \citep{Bovy15}, which is used to model the gravitational potentials and perform the orbital integrations while incorporating the relevant physical processes. It also interfaces with the publicly available \texttt{AGAMA} package\footnote{\url{https://github.com/GalacticDynamics-Oxford/Agama}} \citep{AGAMA}, which is employed to generate equilibrium particle distributions when required. \texttt{CosmoDyn} has already been used in several studies \citep{Chu23,Boldrini25,Boldrini26,Adrian26}, and is currently being applied to an ongoing Euclid Collaboration project dedicated to the dynamical evolution of GCs (Boldrini et al., in preparation).

The target systems are selected through \texttt{GALAXY\_IDS}, which contains a list of galaxy identifiers consistent with the adopted cosmological simulation. For large samples, setting \texttt{CONTINUE\_ON\_ERROR=True} allows the pipeline to continue with the remaining galaxies if the calculation fails for one particular system. In the current implementation, users must prepare the input files for each galaxy listed in \texttt{GALAXY\_IDS} from their cosmological simulation before running \texttt{CosmoDyn}; these files are not generated automatically by the code. For the host galaxy, they consist of a formatted text file, \texttt{DataG<ID>.txt}, containing the properties of the main progenitor at each simulation snapshot, and a serialized file, \texttt{PotsG<ID>.pkl}, containing the corresponding analytical \texttt{galpy} potentials. Equivalent files must be prepared for the progenitors of the satellite galaxies. \texttt{CosmoDyn} then reads these user-provided files to reconstruct the time-dependent galactic environment and perform the subsequent orbital integrations. The exact file formats, required quantities, column definitions, and adopted units are provided in Appendix~\ref{ApA}. Although the extraction of galaxy properties from cosmological simulations is not performed by \texttt{CosmoDyn}, Appendix~\ref{ApB} provides a practical example showing how simulation-extracted data can be converted into the input files required by the code.

The astrophysical systems that \texttt{CosmoDyn} is designed to follow are objects whose gravitational influence on their host galaxy remains negligible, allowing them to be reasonably approximated as test particles or point masses. These include GCs, NSCs, BHs and single stars. Throughout this paper, we collectively refer to these systems as \emph{objects}. Once the evolving galactic potentials have been reconstructed, \texttt{CosmoDyn} provides a library of time-dependent galactic potentials within which the orbits of these objects can be integrated while accounting for the dominant physical processes governing their evolution, such as dynamical friction and mass loss. 

The execution of \texttt{CosmoDyn} is organized into successive pipeline stages controlled through \texttt{RUN\_MODE}. The predefined modes \texttt{RUN\_MODE="in\_situ"} and \texttt{RUN\_MODE="ex\_situ"} activate the stages required to follow objects formed in the main host galaxy and in its satellites, respectively, while \texttt{RUN\_MODE="full"} combines both formation channels. In these three modes, the required stages are selected automatically. Alternatively, \texttt{RUN\_MODE="custom"} allows each stage to be activated or disabled individually through the corresponding parameters. This mode is particularly useful for reusing previously generated files without repeating the complete calculation. \texttt{CosmoDyn} is an open-source Python package publicly available at \url{https://github.com/Blackholan/CosmoDyn}. A complete commented launcher and a step-by-step getting-started guide are provided in 
Appendice~\ref{ApB}, respectively.

In the following, we provide a technical overview of the main steps of the \texttt{CosmoDyn} pipeline. We first reconstruct the time-dependent gravitational potential of the main host galaxy and those of its satellites. We then define the populations of compact objects to be followed and generate their initial conditions within the host or satellite galaxies. Finally, we integrate their orbits while accounting for the selected physical processes and, when requested, model the formation and evolution of their tidal streams.

\subsection{Time-dependent host galaxy potential}\label{HostSection}

Our approach reconstructs the time evolution of a galaxy's gravitational potential from its main progenitor branch extracted from a cosmological simulation. The beginning of the calculation is selected through \texttt{SNAPSHOT\_INDEX}, which identifies the first simulation snapshot included in the reconstruction. The final epoch is controlled by \texttt{END\_SNAPSHOT\_INDEX}. Setting \texttt{END\_SNAPSHOT\_INDEX=None} follows the galaxy to the last available snapshot, which generally corresponds to $z=0$, whereas an integer value stops the reconstruction at the corresponding earlier snapshot. In the application presented below, we use \texttt{SNAPSHOT\_INDEX=0} and \texttt{END\_SNAPSHOT\_INDEX=None} to follow the selected main progenitor from $z=3$ to the present day.

\subsubsection{Spherical potentials}\label{sec:spherical_potentials}

Unlike approaches relying directly on the simulation particles, \texttt{CosmoDyn} reconstructs a continuous analytical representation of each galaxy from its global structural properties. At every snapshot, the stellar and DM masses, half-mass radii, together with the position and velocity of the galaxy centre, are extracted from the simulation and used to construct an analytical gravitational potential. The stellar component is modelled with a Hernquist profile \citep{Hernquist90}, while the DM halo is described by a Navarro--Frenk--White (NFW) profile \citep{Navarro97},

\begin{equation}
\rho_{\mathrm{DM}}(r)=
\frac{\rho_0}
{\left(\dfrac{r}{r_s}\right)
\left(1+\dfrac{r}{r_s}\right)^2},
\end{equation}
where $r$ is the distance from the centre of the DM halo, and $\rho_0$ and $r_s$ denote the characteristic density and scale radius, respectively. 

These choices define the default potential-construction interface used for the application presented in this work. The reconstructed components should be stored in \texttt{PotsG<ID>.pkl} and are subsequently loaded by the dynamical pipeline. Consequently, changing the density profile requires rebuilding this potential file, but does not require modifying the initial-condition or orbital-integration stages. This separation allows the same \texttt{CosmoDyn} workflow to be applied to different DM profiles or baryonic decompositions using any potential implemented in \texttt{galpy}. Although the NFW profile is the standard prediction of $\Lambda$CDM-only simulations, stellar or BH feedback, as well as alternative DM scenarios, may substantially modify the inner DM distribution. In particular, numerous observational studies have suggested that some galaxies host central cores with nearly constant-density inner profiles rather than cuspy NFW profiles \citep{2001AJ....122.2381M,2015AJ....149..180O,Oman15,2020MNRAS.495...58S}. To account for such possibilities, \texttt{CosmoDyn} can readily employ alternative analytical density profiles via \texttt{TwoPowerSphericalPotential}. 

In particular, the current implementation includes a dedicated prescription for FDM haloes. In this framework, DM is modelled as a non-relativistic, ultra-light scalar field with negligible self-interactions \citep{2000NewA....5..103G,2000PhRvL..85.1158H}. The constituent particles have extremely low masses, typically in the range $m_\chi=0.1$--$100\times10^{-22}$~eV, giving rise to quantum mechanical effects on astrophysical scales. FDM haloes are expected to consist of a central solitonic core surrounded by an outer NFW-like envelope \citep{Schive14}. To reproduce this structure, we approximate the halo as the sum of two \texttt{TwoPowerSphericalPotential} components available in \texttt{galpy}. The central solitonic core is represented by
\begin{equation}
\rho_{\rm DM}(r)=
\frac{\rho_{01}}{4\pi r_{c1}^3}
\left[1+\left(\frac{r}{r_{c1}}\right)\right]^{-\beta},
\label{galpy1}
\end{equation}
while the outer halo is approximated by
\begin{equation}
\rho_{\rm DM}(r)=
\frac{\rho_{02}}{4\pi r_{c2}^3}
\left[1+\left(\frac{r}{r_{c2}}\right)\right]^{-3},
\label{galpy2}
\end{equation}
which reproduces the pseudo-NFW envelope outside the core region. The parameters $(\rho_{01},\rho_{02},r_{c1},r_{c2})$ are determined by minimizing the quadratic error with respect to the target soliton-plus-envelope density profile. This approximation reproduces the expected transition between the solitonic core and the outer halo, including the logarithmic slope of $-1$ predicted by FDM simulations \citep{Boldrini26}. Expressing the FDM halo as a combination of existing \texttt{galpy} potentials provides an excellent compromise between computational efficiency and accuracy, while avoiding the implementation of a dedicated potential that would require further optimization within \texttt{galpy}. The choice of the gravitational potential and that of the dynamical 
friction prescription are treated independently. In particular, the FDM potential described above has to be defined during the potential-construction stage, whereas the parameter \texttt{DF\_MODEL}, introduced in Section~\ref{sec:df}, controls the drag force applied during the orbital integration.

\subsubsection{Non-spherical potentials}

The previous models assume spherical symmetry. However, \texttt{CosmoDyn} also supports non-spherical DM haloes through the \texttt{TwoPowerTriaxialPotential} implemented in \texttt{galpy}. The corresponding density profile is given by
\begin{equation}
\rho_{\rm DM}(x,y,z)=
\frac{\rho_0}
{\left(m/r_s\right)^{\alpha}
\left(1+m/r_s\right)^{\beta-\alpha}},
\label{eq:triaxial_density}
\end{equation}
where the ellipsoidal radius $m$ is defined as
\begin{equation}
m^2=x^2+\frac{y^2}{b^2}+\frac{z^2}{c^2}.
\label{eq:ellipsoidal_radius}
\end{equation}
The parameters $\alpha$ and $\beta$ determine the inner and outer logarithmic slopes of the density profile, respectively. In particular, $\alpha=1$ and $\beta=3$ recover an NFW-like profile in the spherical limit. The axis ratios $b$ and $c$ characterize the halo shape, with $b=c=1$ corresponding to a spherical halo, while departures from unity describe increasingly flattened or triaxial configurations. Cosmological $\Lambda$CDM simulations predict that DM haloes are generically triaxial, with typical minor-to-major axis ratios of $c/a\simeq0.5$ \citep{Allgood06}. MW-like haloes are expected to be only moderately flattened, with typical axis ratios of $b/a\simeq0.8$ and $c/a\simeq0.7$ \citep{Jing02}.

\subsection{Interpolation}

The analytical potentials reconstructed for \texttt{CosmoDyn} provide a continuous description of the time evolution of the host galaxy throughout cosmic history. Cosmological simulations provide galaxy properties only at discrete snapshots. To prevent artificial fluctuations arising from the finite temporal resolution and numerical noise of cosmological simulations, the structural parameters defining each analytical component, such as its mass and characteristic radius, are linearly interpolated between consecutive snapshots. The interpolation is therefore applied to the parameters of the potential rather than directly to the potential. Such interpolation is particularly important for simulations with a limited number of snapshots, where abrupt changes between successive outputs could otherwise introduce spurious orbital perturbations. This operation is not carried out by \texttt{CosmoDyn}, which reads the resulting sequence of prepared potentials and uses it to describe the continuous evolution of the galactic environment.

\subsection{Time-dependent satellite galaxy potentials}

We next identify the satellite galaxies accreted by the host galaxy throughout its assembly history. To do so, we traverse the snapshots of the host galaxy main progenitor branch and, at each epoch, identify the galaxies whose descendant corresponds to the host galaxy main progenitor at the subsequent snapshot. The selected galaxies are therefore those that merge with the main branch between two consecutive snapshots. To ensure that the structural properties of the satellites are sufficiently well resolved, we retain only galaxies with stellar masses larger than 100 times the stellar mass resolution of the simulation. This criterion guarantees robust estimates of their structural parameters. For each selected satellite, we reconstruct its own main progenitor branch back to the initial redshift considered or, if applicable, to the earliest snapshot available before its accretion. At every snapshot, we extract the stellar and DM masses, characteristic radii, as well as the position and velocity of the satellite. Positions are converted into physical coordinates, while velocities are corrected for the appropriate cosmological factor. As described in Section~\ref{HostSection}, these structural properties are then used to construct an evolving analytical potential consisting of an NFW DM halo and a Hernquist stellar component. Each satellite is therefore represented by its own time-dependent analytical potential. The integration of each satellite is limited through \texttt{MAXIMUM\_SATELLITE\_RADIUS} in kpc. At the beginning of each snapshot interval, \texttt{CosmoDyn} evaluates the current galactocentric radius of the satellite. If this radius is equal to or larger than \texttt{MAXIMUM\_SATELLITE\_RADIUS}, the integration of that satellite is stopped. 

The satellite galaxies are subsequently integrated within the time-dependent potential of the host galaxy in order to reconstruct their orbital evolution during the accretion process. Although their trajectories are directly available from the cosmological simulation, we choose to re-integrate their orbits for two main reasons. First, the temporal resolution of cosmological simulations is often limited by the irregular spacing between snapshots, whereas orbital integration provides a much finer sampling of the trajectories. Second, this approach enables satellites to be represented as self-consistent moving analytical potentials that can interact with the \emph{objects} throughout the orbital integration. To account for the orbital decay of massive satellites, we include dynamical friction using the Chandrasekhar formulation implemented in \texttt{galpy} (see Section~\ref{DFC}). The satellite mass is taken as the sum of its stellar and DM components, while an effective radius is estimated from their respective characteristic radii. This treatment reproduces the gradual loss of orbital energy and the inward migration of satellites within the host galaxy potential. Applying this framework to approximately 200 MW-like galaxies from TNG50 \citep{Boldrini25}, we showed that our model slightly overestimates the strength of dynamical friction during the early stages of satellite evolution, leading to a somewhat faster orbital decay than observed in the cosmological simulation. These differences mainly arise from the simplifying assumptions of our analytical framework, which models both the host galaxy and the satellites with smooth spherical density distributions and describes dynamical friction through an analytical prescription (see Appendix~A of \citealt{Boldrini25}). In contrast, the gravitational field in TNG50 is intrinsically non-spherical and contains substructures, asymmetries, and complex local gravitational interactions. Moreover, dynamical friction naturally emerges from the self-consistent gravitational interactions between particles rather than being imposed analytically.

The reconstruction of the satellite trajectories constitutes a separate pipeline stage. In the predefined \texttt{RUN\_MODE="ex\_situ"} and \texttt{RUN\_MODE="full"} configurations, this stage is activated automatically. With \texttt{RUN\_MODE="custom"}, setting \texttt{RUN\_EX\_SITU\_SATELLITES=True} integrates the satellite orbits and constructs their time-dependent analytical potentials.

\subsection{Construction of the full cosmological potential}

Each satellite is represented by an analytical potential composed of an NFW DM halo and a Hernquist stellar component, whose masses and structural parameters evolve with time according to the properties extracted directly from TNG50. Once its orbit has been computed, the satellite is incorporated into the combined potential as a \texttt{MovingObjectPotential}. The final model therefore simultaneously describes the structural evolution of the host galaxy, the orbital evolution of the accreted satellites, and their gravitational influence on the \emph{objects} throughout the assembly history of the galaxy. The parameter \texttt{INCLUDE\_MOVING\_SATELLITES} determines whether the gravitational fields of the moving satellites are included in the potential used to integrate the compact objects. This option requires the satellite trajectories and the full host-plus-satellites potential 
to have been generated beforehand by setting \texttt{RUN\_EX\_SITU\_SATELLITES=True}. When \texttt{INCLUDE\_MOVING\_SATELLITES=True}, the objects evolve in the combined potential of the host galaxy and its moving satellites. When it is set to \texttt{False}, only the host-galaxy potential is used. Comparing this configuration with \texttt{INCLUDE\_MOVING\_SATELLITES=True} makes it possible to quantify the dynamical impact of the satellites on the integrated objects. This choice does not determine whether the satellite trajectories themselves are integrated. After a satellite merges with the host galaxy (Figure~\ref{fig2}), its individual contribution is removed from the combined potential, leaving only the evolution of the host galaxy. The time dependence of the host potential is controlled through \texttt{POTENTIAL\_MODE}. With \texttt{POTENTIAL\_MODE="evolving"}, the pipeline follows the complete sequence of analytical potentials between \texttt{SNAPSHOT\_INDEX} and \texttt{END\_SNAPSHOT\_INDEX}. With \texttt{POTENTIAL\_MODE="static"}, a single potential is kept fixed throughout the integration. In the latter case, the potential is selected through \texttt{STATIC\_POTENTIAL\_INDEX}; this parameter is ignored when \texttt{POTENTIAL\_MODE="evolving"}. 

The resulting time-dependent potential is then discretized for the orbital integrations. For each pair of consecutive snapshots, the corresponding cosmological time interval is divided into an integer number of integration steps, with the timestep specified by the user. In \citet{Boldrini25}, a timestep of $2\,\mathrm{Myr}$ (500 integration steps per Gyr) was found to provide smooth and accurate orbital integrations over the full evolution. The number of steps assigned to each interval is adjusted so that the total integrated time exactly matches the cosmic time elapsed between the first and last snapshots. The same timestep is adopted for integrating both the satellite orbits and the trajectories of the \emph{objects} within the evolving MW environment. The mapping between cosmological times and simulation snapshots is provided through \texttt{TIMESTEP\_FILE}. This text file specifies the successive time intervals and the potential index associated with each interval. It therefore determines when the galactic potential is updated during the orbital calculation. The same temporal discretization is used for the satellite trajectories and for the GC, NSC, and BH integrations, ensuring that all components evolve within a consistent cosmological timeline. The required columns and units of \texttt{TIMESTEP\_FILE} are listed in Appendix~\ref{ApA}. In terms of computational performance, constructing the complete time-dependent potential for approximately 200 MW-like environments requires about 4 CPU hours.

The principal outputs of this stage are the time-ordered host potentials, the trajectories and moving potentials of the selected satellites, and the resulting full galactic potential. These serialized potential files define the cosmological environment independently of the compact-object populations subsequently evolved within it. They therefore need to be constructed only once and can then be reused to explore different initial conditions, add or remove GCs, NSCs, or BHs, and vary the dynamical-friction and mass-loss prescriptions without repeating the host or satellite reconstruction.

\begin{figure}
    \centering
    \includegraphics[width=1\linewidth]{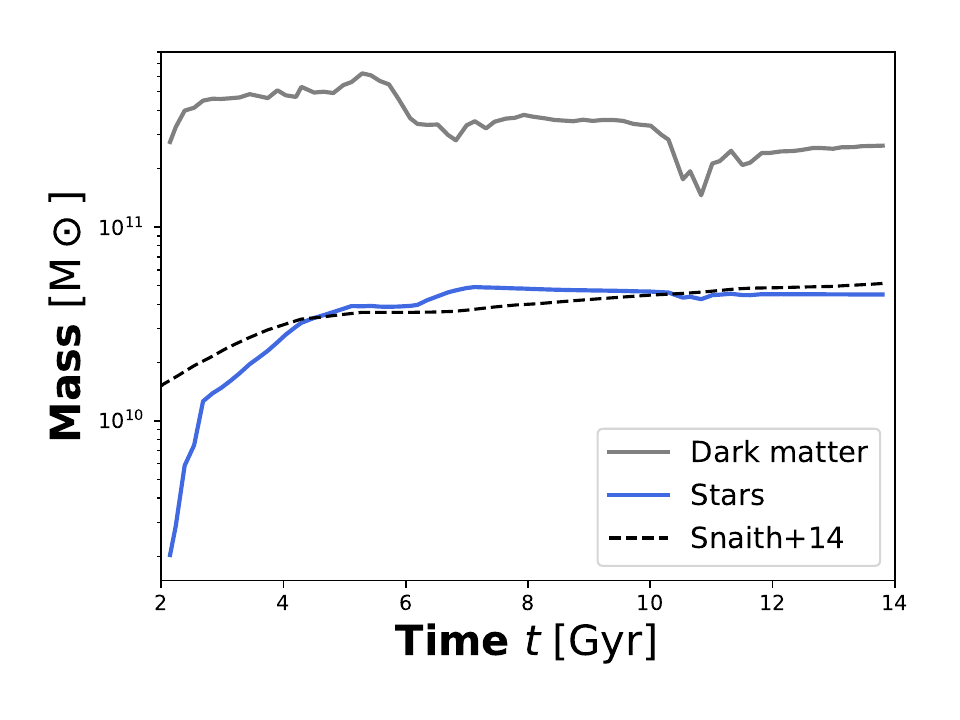}
    \caption{Mass evolution of the stellar and DM components of our reference MW analogue (ID 613192) selected from the sample of 198 MW-like galaxies identified in TNG50. This galaxy provides the closest match to the stellar mass assembly history of the MW inferred by \citet{Snaith2015} over the past $\sim7$ Gyr. At $z=0$, its DM and stellar masses are $M_{\rm DM}=2.6\times10^{11}\,\mathrm{M_\odot}$ and $M_\star=4.5\times10^{10}\,\mathrm{M_\odot}$, respectively. The DM mass loss may be explained by past interactions between our MW analogue and its five neighbouring galaxies identified at $z=0$, whose DM masses are of the order of $10^{11}\,\mathrm{M_\odot}$.}
    \label{fig1}
\end{figure}

\subsection{Illustrative application to a Milky Way-like galaxy}

As an illustration, the methodology described above is applied to our reference MW analogue identified in the TNG50 cosmological hydrodynamical simulation in order to reconstruct the time-dependent gravitational potential of both the host galaxy and its satellite population. TNG50 is a high-resolution cosmological hydrodynamical simulation of a $(51.7\,{\rm Mpc})^3$ volume containing $2\times2160^3$ resolution elements. It reaches a baryonic mass resolution of $8.5\times10^4\,{\rm M_\odot}$ and a DM mass resolution of $4.5\times10^5\,{\rm M_\odot}$, with a gravitational softening length of 288 pc at $z=0$. The simulation includes state-of-the-art prescriptions for star formation, chemical enrichment, BH growth, AGN feedback, and galactic winds \citep{2017MNRAS.465.3291W,2018MNRAS.473.4077P}. Using TNG50, \citet{Pillepich24} identified 198 MW-like galaxies whose stellar masses ($10^{10.5}<M_\star/{\rm M_\odot}<10^{11.2}$) and large-scale environments resemble those of the MW. Among this sample, we selected as an example the galaxy with subhalo ID 613192, as it provides the closest agreement with the reconstructed stellar mass assembly history of the MW derived by \citet{Snaith2015} over the past $\sim7$ Gyr (Figure~\ref{fig1}). The main progenitor branch of this galaxy is followed from $z=3$ to $z=0$, corresponding to 75 simulation snapshots.

\begin{figure}
    \centering
    \includegraphics[width=1\linewidth]{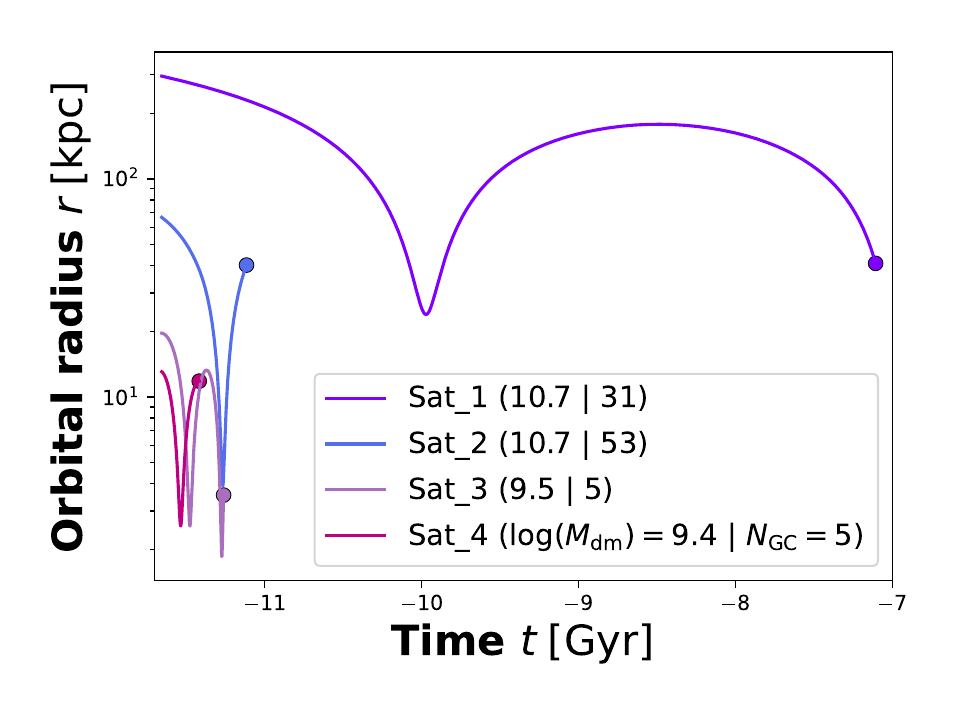}
    \caption{Orbital radius as a function of time for the four satellite galaxies merging with our reference MW analogue. The filled circles indicate the merger times measured in the TNG50 simulation.}
    \label{fig2}
\end{figure}

The satellite galaxies accreted by this MW analogue are identified as subhaloes with stellar masses $M_\star>10^7\,{\rm M_\odot}$ whose descendants merge with the main progenitor branch. This threshold corresponds to more than one hundred stellar particles in TNG50, ensuring that the structural properties of the satellites are sufficiently well resolved. For each satellite, we store the time evolution of its stellar and DM masses, characteristic radii, position, and velocity in order to reconstruct its own evolving analytical potential prior to accretion. Figure~\ref{fig2} presents the re-integrated orbits of the four most massive satellites accreted by our reference MW analogue. The orbital integrations account for the time evolution of the satellite masses extracted from the cosmological simulation.

\section{Initial conditions of compact stellar systems and black holes}\label{sec:initial_conditions}

\begin{figure}
    \centering
    \includegraphics[width=1\linewidth]{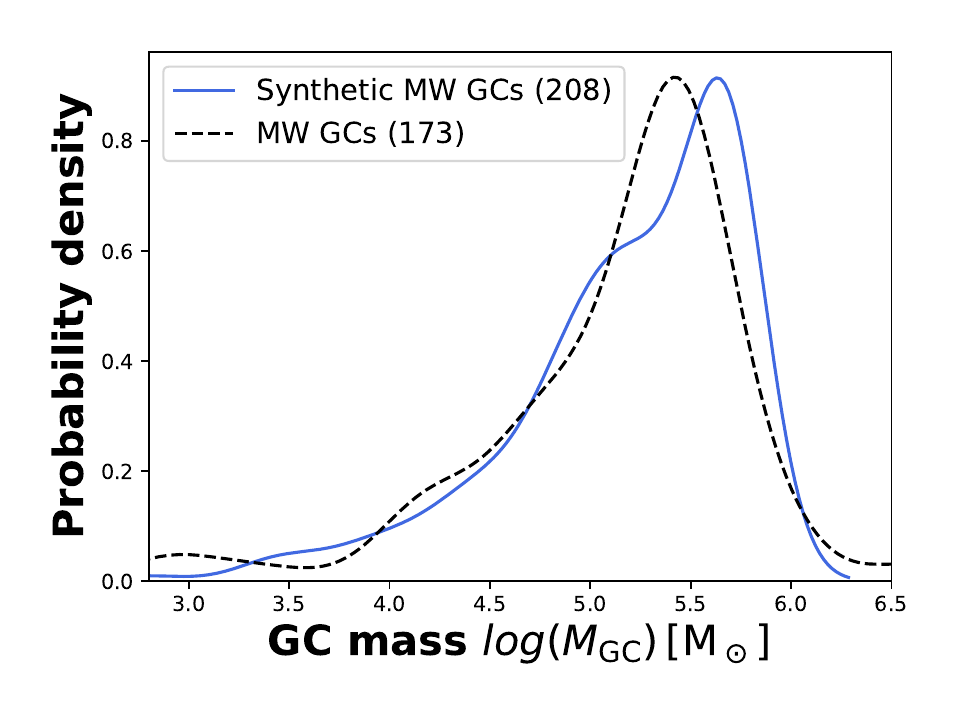}
    \caption{Probability density distribution of GC masses at $z=0$ for the synthetic GC population of our reference MW analogue. The black dashed curve represents the observed MW GC mass distribution \citep{2021MNRAS.505.5978V}, illustrating the good agreement achieved by our calibrated model.}
    \label{fig3}
\end{figure}

After reconstructing the time-dependent gravitational potential of the host galaxy and its satellite population, the next step is to define the initial conditions of the \emph{objects} whose dynamical evolution will be followed within this evolving cosmological environment. \texttt{CosmoDyn} distinguishes two populations according to their formation site: \textit{in-situ} \emph{objects}, formed within the host galaxy, and \textit{ex-situ} \emph{objects}, formed in satellite galaxies before being accreted by the host. Both formation channels are available for GCs, NSCs, and BHs and may be activated independently 
when \texttt{RUN\_MODE="custom"} is used. In the predefined execution modes, the formation channels are selected automatically. The mode \texttt{RUN\_MODE="in\_situ"} generates only objects associated with the main host galaxy, \texttt{RUN\_MODE="ex\_situ"} generates only objects associated with the satellite galaxies, and \texttt{RUN\_MODE="full"} combines both populations. The generation of GCs is part of the standard pipeline, whereas NSCs and BHs are optional populations activated through \texttt{ENABLE\_NSCS=True} and \texttt{ENABLE\_BHS=True}, respectively. Setting either parameter to \texttt{False} removes the corresponding population from the calculation. With \texttt{RUN\_MODE="custom"}, the initial-condition stages are controlled individually through \texttt{RUN\_ICS} and 
\texttt{RUN\_EX\_SITU\_ICS} for GCs, \texttt{RUN\_IN\_SITU\_NSC\_ICS} and \texttt{RUN\_EX\_SITU\_NSC\_ICS} for NSCs, and 
\texttt{RUN\_IN\_SITU\_BH\_ICS} and \texttt{RUN\_EX\_SITU\_BH\_ICS} for BHs. This allows an existing set of initial conditions to be reused without regenerating it before each dynamical calculation. In the application presented in this work, we consider both the 
\textit{in-situ} and \textit{ex-situ} GC populations, but restrict the BH and NSC samples to \textit{ex-situ} objects. Their \textit{in-situ} counterparts are assumed to correspond to the central BH and NSC already residing at the centre of the MW analogue and are therefore not followed.

The initial positions and velocities of the objects can either be provided directly by the user or generated by \texttt{CosmoDyn}. GC phase-space coordinates are sampled from equilibrium distribution functions constructed with \texttt{AGAMA}, whereas the NSCs and BHs generated by the standard pipeline are placed on circular orbits in the centre of their host galaxy. \textit{In situ} coordinates are defined relative to the main host, while \textit{ex situ} coordinates are defined in the reference frame of the corresponding satellite. The exact format and units of the initial-condition files are provided in Appendix~\ref{ApA}. This framework provides a self-consistent description of both native and accreted populations while naturally accommodating additional physical processes, such as dynamical friction, mass loss, and satellite accretion.

\subsection{Initial conditions for globular clusters}

Here we describe how the initial GC population is assigned to each galaxy at a given redshift. The number of GCs is determined from the empirical relation of \citet{Burkert20}, which links the total GC population to the virial mass of the host halo:
\begin{equation}
\langle \log N_{\rm GC} \rangle =
-9.58 \pm 1.58 +
(0.99 \pm 0.13)
\log \left(\frac{M_{\rm vir}}{M_\odot}\right).
\label{eq:ngc_relation}
\end{equation}
We included uncertainties as random noise in our modelling. Since this relation is calibrated using nearby galaxies, its applicability at high redshift ($z\simeq2$--3) remains uncertain. \citet{2019MNRAS.488.5409C} suggest that the normalization of the GC--halo mass relation may evolve by up to an order of magnitude between $z=3$ and $z=0$. For this reason, \texttt{CosmoDyn} allows the normalization of the relation to be multiplied by a user-defined scaling factor, \texttt{ALPHA}. For the in-situ population, the number of GCs is controlled through \texttt{NGC}. Setting \texttt{NGC} to a strictly positive integer directly imposes the requested number of clusters. Setting \texttt{NGC=0} instead computes the number of GCs from Equation~\ref{eq:ngc_relation}. In the latter case, the predicted number is multiplied by the user-defined normalization factor, \texttt{ALPHA}. Consequently, \texttt{ALPHA} is ignored whenever \texttt{NGC>0}. The corresponding parameters for the ex-situ population are \texttt{NGC\_EX\_SITU} and \texttt{ALPHA\_EX\_SITU}. They follow the same convention but are applied independently to each selected satellite galaxy. In our application, setting we applied the same number prescription to the host and satellite populations.

The initial phase-space distribution of \textit{in-situ} GCs is generated from equilibrium distribution functions constructed with \texttt{AGAMA}\footnote{\url{https://github.com/GalacticDynamics-Oxford/Agama}}. For each galaxy, we build an equilibrium model consisting of a stellar component embedded in a DM halo using the structural properties extracted from the cosmological simulation at the chosen redshift. For each galaxy, the analytical stellar and DM potentials used to describe its gravitational field are also supplied to \texttt{AGAMA} at the selected initial snapshot. \texttt{AGAMA} uses these potentials to construct an equilibrium distribution function. We then sample the resulting distribution function to generate the initial positions and velocities of the stellar particles, which are used as proxies for the GC population. In the current implementation, all GCs are initialized at the same chosen redshift. If a satellite galaxy has not yet formed at this epoch, its GC population is instead initialized at the earliest snapshot in which the satellite is identified. 

The equilibrium distribution is sampled with \texttt{N\_PARTICLES\_PER\_COMPONENT} particles per galactic component. The spatial region sampled by \texttt{AGAMA} is controlled through \texttt{TAGGING\_RADIUS\_FACTOR}, which defines the tagging radius relative to the stellar half-mass radius of the galaxy. The parameter \texttt{MINIMUM\_TAGGING\_RADIUS}, in kpc, imposes a minimum distance for the GC population. Setting \texttt{MINIMUM\_TAGGING\_RADIUS=None} disables this lower limit. The orbital selection of the GC candidates is controlled through \texttt{CIRCULARITY\_THRESHOLD}. When this parameter is assigned a numerical value $\epsilon_{\rm min}$, only particles satisfying $L_z \geq \epsilon_{\rm min}L_{\rm circ}(E)$ are retained, where $L_{\rm circ}(E)$ is the angular momentum of a circular orbit with the same orbital energy. For example, \texttt{CIRCULARITY\_THRESHOLD=0.6} selects preferentially prograde, disk-like orbits. Setting \texttt{CIRCULARITY\_THRESHOLD=None} disables the circularity selection and allows the GCs to be drawn from the full spatially selected particle population.

All GCs are assigned the initial mass \texttt{GC\_MASS}, in $M_\odot$, and the half-mass radius \texttt{GC\_HALF\_MASS\_RADIUS}, in kpc. These quantities are used during the dynamical stage. The half-mass radius remains fixed, whereas the mass may evolve according 
to the prescription selected through \texttt{MASS\_LOSS\_MODE} (see Section~\ref{MLP}).

\subsection{Initial conditions for black holes and nuclear star clusters}

NSCs and BHs are optional populations controlled through \texttt{ENABLE\_NSCS} and \texttt{ENABLE\_BHS}, respectively. When 
enabled, \texttt{CosmoDyn} can generate one in-situ object associated with the main host galaxy and one ex-situ object for each selected satellite. In our application where the central BH or NSC of the main host is not of interest, we assign one BH and one NSC to each accreted satellite galaxy. This is the configuration adopted below when we focus only on compact remnants delivered by accreted satellites, which requires us to use \texttt{RUN\_MODE="custom"}. For the reference MW analogue shown in Figure~\ref{fig2}. This results in four BHs and four NSCs due to the accretion of four 
satellite galaxies, whose properties are discussed in Section~\ref{MLP}.

The BH mass is controlled through \texttt{BH\_MASS}. A strictly positive value assigns the same user-defined mass to every in-situ and ex-situ BH. Alternatively, setting \texttt{BH\_MASS=0} computes the mass of each BH from the stellar mass of its host galaxy using the empirical relation of \citet{1998AJ....115.2285M}:
\begin{equation}
M_{\rm BH}=0.006\,M_\star,
\label{eq:bh_mass}
\end{equation}
where $M_\star$ is evaluated at the redshift at which the BH is initialized. For the galaxies considered in this work, this prescription yields BH masses of approximately $10^6$--$10^7,\mathrm{M_\odot}$. For an in-situ BH, $M_\star$ is the stellar mass of the main host at \texttt{SNAPSHOT\_INDEX}; for an ex-situ BH, it is the stellar mass of the parent satellite at its tagging snapshot. BHs are always treated as point masses: their half-mass radius is fixed internally to zero and is therefore not exposed as a free parameter. Their masses remain constant during the integration, independently of the value selected through \texttt{MASS\_LOSS\_MODE} (see Section~\ref{MLP}).

The NSC mass is estimated using the scaling relation of \citet{Nadine20}, including its intrinsic scatter of 0.6 dex:
\begin{equation}
\log \left(M_{\rm NSC}/M_\odot\right)
=
0.48\,\log\left(\frac{M_\star}{10^9\,M_\odot}\right)
+6.51.
\label{eq:nsc_mass}
\end{equation}
This relation predicts NSC masses of approximately $10^6\,\mathrm{M_\odot}$ for the galaxies considered in this work. As for BHs, the NSC mass is computed from the scaling relation in Eq.~\eqref{eq:nsc_mass} when \texttt{NSC\_MASS=0}; otherwise, the positive value specified by \texttt{NSC\_MASS} is assigned to all NSCs. They are assigned the half-mass radius \texttt{NSC\_HALF\_MASS\_RADIUS}, in kpc. Unlike BHs, NSCs are treated as extended stellar systems and may undergo mass loss during their dynamical evolution. As for GCs, NSCs are assigned a fixed half-mass radius of 10 pc throughout the integrations. The \texttt{NSC\_HALF\_MASS\_RADIUS} parameter has a negligible impact on their evolution, as dynamical friction is negligible within the MW halo, while inside the parent satellite galaxy the NSC is already located at the dynamical centre. The initial galactocentric radii of the NSCs and BHs are controlled through \texttt{NSC\_INITIAL\_RADIUS} and \texttt{BH\_INITIAL\_RADIUS}, respectively, in kpc. Each object is initialized on a circular orbit at the selected radius in the host potential at \texttt{SNAPSHOT\_INDEX}. For ex-situ objects, the same calculation is performed in the reference frame and gravitational potential of the parent satellite. In the application presented in this work, we adopt an initial radius of 1 pc, as they should lie in the central region of the galaxy. We verified that adopting a smaller initial radius does not significantly modify the results.

\subsection{Initial globular cluster population of the reference Milky Way}

We generate the GC population at $z=3$ (\texttt{SNAPSHOT\_INDEX=0}), assigning \textit{in-situ} GCs to the progenitor of the MW and \textit{ex-situ} GCs to the progenitors of the merging satellite galaxies. We further require all $N_{\rm GC}$ clusters to be initially located within the stellar half-mass radius of their host galaxy and to satisfy $L_z > 0.6\,L_{\rm circ}$ (\texttt{CIRCULARITY\_THRESHOLD=0.6}), where $L_{\rm circ}$ is the angular momentum of a circular orbit with the same energy. This selection places the GCs on disk-like orbits, although the analytical gravitational potentials adopted in this work do not explicitly include a stellar disk component. The total number of GCs is estimated from the relation of \citet{Burkert20} (when \texttt{NGC=0} and \texttt{NGC\_EX\_SITU=0}), multiplied by a free normalization factor calibrated to reproduce the present-day MW GC system. The calibration is performed by matching both the total number of surviving GCs and their mass distribution at $z=0$ (Figure~\ref{fig3}). For the reference MW analogue, a normalization factor of three (\texttt{ALPHA=3}) reproduces the observed GC population of the MW with 208 surviving GCs, including 123 \textit{in-situ} clusters \citep{2024A&A...687A.214G}. This factor increases the initial GC population to compensate for the large number of clusters destroyed during their subsequent evolution. Figure~\ref{fig3} shows the probability density distribution of GC masses at $z=0$ with the observed MW GC mass distribution from \citet{2021MNRAS.505.5978V}. All GCs are assigned an initial mass of $10^6\,{\rm M_\odot}$ (\texttt{GC\_MASS=$10^6$}) and a half-mass radius of 10 pc (\texttt{GC\_HALF\_MASS\_RADIUS=0.01}). For simplicity, we currently adopt the same initial mass for all GCs, as this assumption produces a reasonable present-day GC mass distribution (see Figure~\ref{fig3}).

\section{Dynamical evolution in a cosmological environment}\label{sec:dynamics}

This final step describes the orbital evolution of the \emph{objects} under the different dynamical processes acting over cosmic time. Once the initial conditions have been defined, the \emph{objects} are integrated within the time-dependent gravitational potential of their host environment. The integration interval begins at \texttt{SNAPSHOT\_INDEX} and ends at \texttt{END\_SNAPSHOT\_INDEX}. When \texttt{END\_SNAPSHOT\_INDEX=None}, the objects are followed to the last available snapshot. The temporal sampling of each interval is read from \texttt{TIMESTEP\_FILEG<ID>.txt}. In our application, the potential parameters are updated every 100 integration steps, corresponding to $200\,\mathrm{Myr}$, using the properties of the host galaxy and its satellites interpolated between consecutive cosmological snapshots. The final phase-space coordinates obtained at the end of each interval are then adopted as the initial conditions for the subsequent integration. Orbital integrations are performed with the fast C-based \texttt{dop853\_c} integrator implemented in \texttt{galpy}. This explicit Runge--Kutta integrator of order 8(5,3) provides an effective compromise between numerical accuracy and computational cost, making it well suited to the integration of large numbers of orbits in time-dependent potentials. The numerical orbit integrator is selected through \texttt{INTEGRATION\_METHOD="dop853\_c"}, which is the default configuration used in this work. Other integration methods supported by \texttt{galpy} may be selected. 

In the predefined execution modes, the dynamical stages are activated automatically according to the selected populations. When 
\texttt{RUN\_MODE="custom"} is used, the in-situ and ex-situ GC dynamics are controlled through \texttt{RUN\_IN\_SITU\_DYNAMICS} and 
\texttt{RUN\_EX\_SITU\_DYNAMICS}, respectively. The corresponding NSC stages are controlled through  \texttt{RUN\_IN\_SITU\_NSC\_DYNAMICS} and \texttt{RUN\_EX\_SITU\_NSC\_DYNAMICS}, while the BH stages are controlled through \texttt{RUN\_IN\_SITU\_BH\_DYNAMICS} and \texttt{RUN\_EX\_SITU\_BH\_DYNAMICS}. Setting one of these parameters to \texttt{False} skips the corresponding integration and allows an existing dynamical output to be reused by subsequent stages.

\textit{In-situ} \emph{objects} remain integrated either within the potential of the main host galaxy alone (\texttt{INCLUDE\_MOVING\_SATELLITES=False}) or within the full potential including the moving satellites (\texttt{INCLUDE\_MOVING\_SATELLITES=True}). By contrast, \textit{ex-situ} \emph{objects} initially evolve within the potential of their parent satellite and are subsequently transferred to the full galaxy potential when they become tidally stripped. This transition is determined dynamically from their orbital properties relative to the satellite. When the orbital energy of an \emph{object} in the satellite reference frame becomes positive, it is considered to be no longer gravitationally bound to its parent galaxy. Its positions and velocities are then transformed into the galactocentric reference frame of the main host, and its subsequent evolution is followed only within the main galactic potential. The release criterion is controlled through \texttt{RELEASE\_ENERGY\_TOLERANCE}. An ex-situ object is considered unbound from its parent satellite at the first integration step for which its satellite-centric orbital energy satisfies $E_{\rm sat} > \texttt{RELEASE\_ENERGY\_TOLERANCE}$. The default value \texttt{RELEASE\_ENERGY\_TOLERANCE=0.0} therefore corresponds to the standard positive-energy criterion. If an object remains bound when the available satellite history ends, it is automatically released from the final position and velocity of the satellite. A very small fraction of the \textit{ex-situ} population, primarily GCs, may remain unbound from the main host at the end of the integrations. For example, \citet{Boldrini25} found only 33 such GCs among a sample of approximately 18\,000 objects. These unbound \emph{objects} are flagged by \texttt{CosmoDyn} before being excluded from the final sample. Since they have escaped the gravitational potential of the main host galaxy, their subsequent orbital integration is no longer relevant to the system under study and is therefore stopped to avoid unnecessary computational costs. The integration of an object is also stopped when its galactocentric radius becomes smaller than or equal to a prescribed central capture radius. This criterion represents unresolved central processes: within this radius, a GC or NSC may be tidally disrupted or incorporated into the galactic nucleus, while a BH may have reached the galaxy centre and subsequently merge with the central BH. The capture radius is specified through \texttt{CENTRAL\_CAPTURE\_RADIUS} for GCs, \texttt{NSC\_CENTRAL\_CAPTURE\_RADIUS} for NSCs, and \texttt{BH\_CENTRAL\_CAPTURE\_RADIUS} for BHs. Once the criterion is satisfied, the object is flagged as captured and its phase-space trajectory is truncated at the first corresponding integration point.

\texttt{CosmoDyn} includes the main physical processes affecting the orbital evolution of the \emph{objects}, including dynamical friction and mass loss. Each process can be activated or disabled independently, allowing the physical complexity of the model to be adapted to the problem under consideration.

\subsection{Dynamical friction implementation}\label{sec:df}

Dynamical friction alters the orbits of compact stellar systems and BHs as they evolve within their host galaxies. Its efficiency depends primarily on the mass ratio between the infalling \emph{object} and the host galaxy. More massive \emph{objects} moving through denser environments experience a stronger gravitational drag, resulting in a more rapid orbital decay. The characteristic timescale over which DF significantly reduces the orbital apocentre can be approximated as $t_{\mathrm{fric}} \sim \left( M_{\mathrm{gal}}(<r) / M_{\mathrm{obj}} \right) \times t_{\mathrm{dyn}}$, where $M_{\rm gal}(<r)$ is the host galaxy mass enclosed within the orbital radius $r$, $M_{\rm obj}$ is the mass of the infalling \emph{object}, and $t_{\rm dyn}$ is the local dynamical time \citep{2008gady.book.....B,DynamicsAstrophysicsGalaxies}. In MW-like galaxies, the enclosed mass is generally several orders of magnitude larger than the mass of a GC or NSC, making DF ineffective over a Hubble time. In contrast, DF can dominate the orbital evolution of \emph{objects} orbiting within low-mass satellite galaxies prior to their accretion. The efficiency of DF also depends on the spatial extent of the infalling \emph{object}. Extended systems generate a less concentrated gravitational wake than point-like masses of the same total mass, reducing the drag force. Consequently, BHs are expected to experience the strongest DF owing to their large masses and their treatment as point particles ($r_{\rm hm}=0$), whereas NSCs and GCs are less affected because of their finite sizes.

The dynamical-friction prescription is selected through \texttt{DF\_MODEL}. Setting \texttt{DF\_MODEL="none"} disables dynamical friction, while \texttt{DF\_MODEL="cdm"} activates the isotropic Chandrasekhar prescription (see Section~\ref{DFC}). The anisotropic formulation of Binney is selected with \texttt{DF\_MODEL="binney"} (see Section~\ref{DFB}), and \texttt{DF\_MODEL="fdm"} activates the FDM prescription (see Section~\ref{FDM}). For GCs and NSCs, the friction force uses the masses \texttt{GC\_MASS} and \texttt{NSC\_MASS} and the half-mass radii \texttt{GC\_HALF\_MASS\_RADIUS} and \texttt{NSC\_HALF\_MASS\_RADIUS}, respectively. For BHs, the individual mass stored in the initial-condition file is used, while the half-mass radius is fixed internally to zero. When the mass of a GC or NSC reaches zero, no dynamical-friction force is subsequently applied to that object.

\subsubsection{Chandrasekhar formulation for spherical systems}\label{DFC}

In the CDM paradigm, an \emph{object} of mass $M_{\rm obj}$ orbiting at a galactocentric distance $r$ with velocity $v$ experiences dynamical friction. Following the classical formulation of \citet{Chandra43}, the drag force is
\begin{equation}
   F_{\mathrm{CDM}}=-4\pi G^2 M_{\mathrm{obj}}^2 \rho(r) \frac{{v}}{v^3} C_{\mathrm{CDM}}(r,v)\,
\end{equation}
where $\rho(r)$ the local density of the host galaxy. The $C_{\rm CDM}$ coefficient :
\begin{equation}
C_{\rm CDM}(r,v)=\frac{1}{2}\ln\!\left(1+\Lambda^{2}\right)\left[
\operatorname{erf}\!\left(\frac{v}{\sqrt{2}\sigma}\right)-\frac{2v}{\sqrt{2\pi}\sigma}
e^{-\frac{v^{2}}{2\sigma^{2}}}\right],
\end{equation}
and the Coulomb factor $\Lambda$ is expressed as
\begin{equation}
\label{eq:lnLambda}
\Lambda= \frac{r}{\max(r_{\mathrm{hm}},\, b_{\mathrm{min}})},
\end{equation}
The dynamical friction implementation via galpy follows \cite{Petts15}, which introduced a semi-analytic model based on \cite{Chandra43} formalism, incorporating radially varying minimum and maximum impact parameters. The parameter $b_{\mathrm{min}} = G M_{\rm obj} / v^2$ corresponds to the impact parameter leading to a $90^\circ$ deflection, while $r_{\mathrm{hm}}$ is the half-mass radius of the object. $G$ is the gravitational constant. The formalism of \cite{Chandra43} was originally derived under the assumption of an infinite, homogeneous background. Despite this limitation, our implementation accounts for the finite size of objects by incorporating their half-mass radius, $r_{\rm hm}$, into the Coulomb logarithm. In addition, its effectiveness can be understood because dynamical friction in real systems arises from resonant interactions that, when forming an effective continuum, recover Chandrasekhar’s expression \citep{Tremaine84,Weinberg86}. Indeed, \cite{Petts15} implementation has been extensively validated against $N$-body simulations, successfully reproducing the inspiral of GCs in spherical halos (both cuspy and core DM distributions) and the core stalling effect without parameter fine-tuning.

\subsubsection{Binney formulation for axisymmetric systems}\label{DFB}

The previous formulation is valid for approximately spherical systems. In non-spherical environments, such as galactic discs or flattened DM haloes, however, the assumption of an isotropic velocity distribution is no longer valid. In this case, the dynamical friction formalism must be modified to account for the anisotropy of the background velocity dispersion. Assuming an axisymmetric velocity distribution, the dynamical friction force in cylindrical coordinates is given by \cite{Binney77}:
\begin{equation}
\mathbf{F}_{\mathrm{CDM,ani}}
=
-\frac{2\sqrt{2\pi}\,G^2M_{\mathrm{obj}}^2\rho(r)}
{\sigma_R^2\sigma_z}
\sqrt{1-e_v^2}\,
\ln\!\left(1+\Lambda^2\right)
\begin{pmatrix}
v_R B_R\\
v_\phi B_R\\
v_z B_z
\end{pmatrix},
\end{equation}
where $e_v^2=1-(\sigma_z/\sigma_R)^2$ and
\begin{equation}
\begin{aligned}
B_R &=
\int_0^\infty
\frac{\mathcal{E}(q)}
{(1+q)^2(1-e_v^2+q)^{1/2}}
\,\mathrm{d}q,\\
B_z &=
\int_0^\infty
\frac{\mathcal{E}(q)}
{(1+q)(1-e_v^2+q)^{3/2}}
\,\mathrm{d}q,
\end{aligned}
\end{equation}
with
\begin{equation}
\mathcal{E}(q)=
\exp\!\left[
-\frac{v_R^2+v_\phi^2}{2\sigma_R^2(1+q)}
-\frac{v_z^2}{2\sigma_R^2(1-e_v^2+q)}
\right].
\end{equation}
The anisotropic dynamical friction formalism adopted here is restricted to axisymmetric systems. In the present implementation, this corresponds to halos with either ($b<1$, $c=1$) or ($b=1$, $c<1$), while the fully triaxial case ($b<1$, $c<1$) is not currently supported. Extending this treatment to fully triaxial haloes would require a more general formulation based on the full three-dimensional velocity dispersion tensor, with independent velocity dispersions ($\sigma_R$, $\sigma_\phi$, and $\sigma_z$) and their possible cross terms.

\subsubsection{Dynamical friction in fuzzy dark matter}\label{FDM}

In the FDM paradigm, the ultra-light mass of the DM particles modifies the dynamical friction experienced by massive \emph{objects}. The wave nature of FDM generates constructive and destructive interference patterns, producing persistent density fluctuations that induce stochastic dynamical heating. This heating counteracts orbital decay, thereby reducing the efficiency of dynamical friction and increasing the infall timescale of objects \citep{Hui17,Lancaster20,BO19}. The FDM dynamical friction force can be approximated as
\begin{equation}
    F_{\rm FDM} = -4\pi G^2 M_{\rm GC}^2 \rho(r)\,\frac{{v}}{v^3}\,C_{\rm FDM}(r,v,m_{22},M_\sigma),
\end{equation}
where the dimensionless coefficient $C_{\rm FDM}$ accounts for the wave properties of FDM. It depends on two additional dimensionless parameters: (i) $m_{22}=m_\chi/10^{-22}\,\mathrm{eV}$, where $m_\chi$ is the FDM particle mass, and (ii) the Mach number $M_\sigma=v/\sigma$, where $\sigma$ is the local velocity dispersion of the host galaxy. This coefficient is the only term that differs from the classical Chandrasekhar formulation. For $m_{22}\gtrsim30$, the FDM dynamical friction converges towards the classical CDM regime, whereas for $m_{22}\lesssim3$, the wave nature of the DM particles substantially suppresses dynamical friction for GCs \citep{Adrian26}. A detailed derivation of $C_{\rm FDM}$ and its implementation in \texttt{CosmoDyn} is presented in \citet{Adrian26}.  The dimensionless particle-mass parameter is then specified through \texttt{M22}. The value of \texttt{M22} must be strictly positive and is ignored for other dynamical friction models. The choice \texttt{DF\_MODEL="fdm"} controls the dynamical-friction 
force and does not, by itself, construct the FDM density profile described in Section~\ref{sec:spherical_potentials}. The corresponding 
FDM gravitational potential must therefore be prepared separately in \texttt{PotsG<ID>.pkl}.

Figure~\ref{fig4} illustrates the deviations from the classical CDM dynamical friction model for a $10^6\,M_\odot$ GC with a half-mass radius of 10~pc, initially located at a galactocentric distance of 2.5~kpc in a dwarf galaxy with a halo mass of $10^9\,M_\odot$. In an axisymmetric halo ($b=0.6$, $c=1$), the combined effects of the anisotropic dynamical friction and the non-spherical gravitational potential slow the orbital decay compared to the spherical CDM case (blue curve in  Figure~\ref{fig4}). In an FDM halo with $m_{22}=1$, the reduction in dynamical friction is even more pronounced, preventing the GC from reaching the galaxy centre within 8 Gyr (orange curve in  Figure~\ref{fig4}). These examples demonstrate that, whenever dynamical friction is efficient, particularly in dwarf galaxies, departures from the classical CDM prescription can significantly alter the orbital evolution of our \emph{objects}. This is especially relevant for GCs later accreted by the MW, whose orbital evolution is largely governed by the dynamical friction they experience while orbiting within their dense progenitor galaxies prior to accretion.

\subsection{Mass-loss prescriptions}\label{MLP}
 
For GCs and NSCs, \texttt{CosmoDyn} accounts for mass loss during the orbital integrations, whereas BHs are assumed to retain a constant mass over time. At each update of the galaxy potential, the mass evolution of compact stellar systems is computed following the prescription of \citet{Kruijssen11}, which includes the effects of stellar evolution, two-body relaxation, and tidal shocks. Mass loss affects the orbital evolution only when dynamical friction is significant, since the frictional force depends explicitly on the instantaneous mass of the \emph{object}. In this case, the mass is updated simultaneously with the galaxy potential so that the dynamical friction force is consistently recomputed throughout the integration. Conversely, when dynamical friction is negligible, as is typically the case for GCs orbiting in MW-like galaxies, mass loss has no measurable impact on the orbital trajectories. In such situations, the orbital integration is first performed assuming a constant mass, and the mass evolution is computed in post-processing to reduce the computational cost. This technique is therefore adopted for the \textit{in-situ} GC population.

Mass loss is controlled through \texttt{MASS\_LOSS\_MODE}, which accepts three values. Setting \texttt{MASS\_LOSS\_MODE="none"} keeps the object mass constant throughout the calculation. With \texttt{MASS\_LOSS\_MODE="postprocess"}, the orbit is first integrated at constant mass and the complete mass history is subsequently computed along the resulting trajectory. The evolving mass therefore does not modify the dynamical-friction force in this mode. Finally, with \texttt{MASS\_LOSS\_MODE="coupled"}, the mass is updated after each snapshot interval and the new value is used to compute dynamical friction during the following interval. The \texttt{"coupled"} mode is required when dynamical friction is efficient, particularly for massive objects orbiting inside low-mass satellite galaxies. The selected value of \texttt{MASS\_LOSS\_MODE} is applied to GCs and NSCs. BHs are assumed to retain a constant mass: for \texttt{OBJECT=BH}, \texttt{CosmoDyn} internally enforces \texttt{MASS\_LOSS\_MODE="none"}, independently of the value specified in the launcher.

\begin{figure}
    \centering
    \includegraphics[width=1\linewidth]{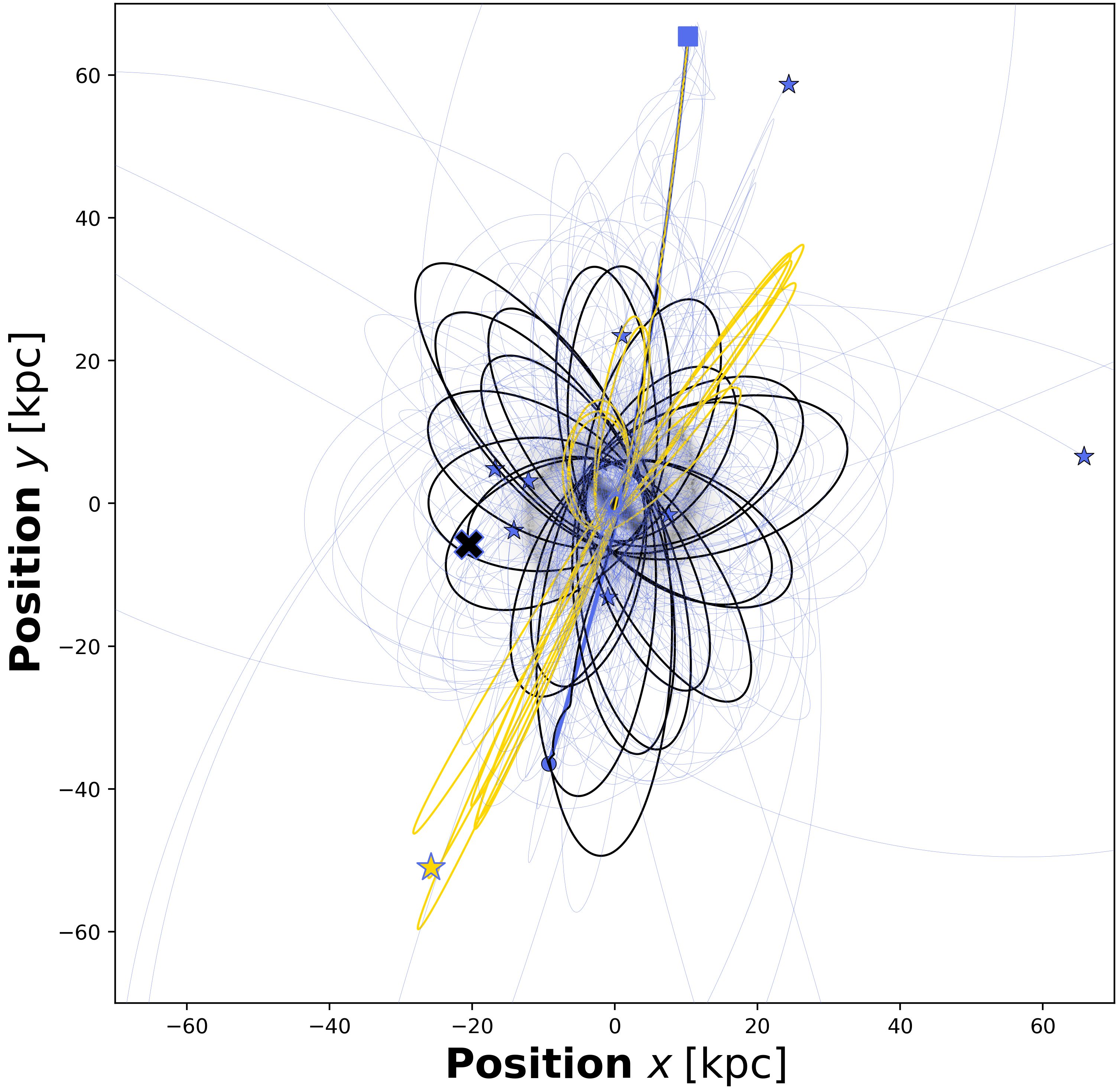}
    \caption{Orbital evolution of the GCs, NSC, and BH associated with satellite 2 of our best MW-like galaxy in the $x$--$y$ plane. Blue stars indicate the present-day ($z=0$) positions of the 8 surviving GCs (out of the 52 initially associated with satellite 2), while the blue curves show their orbital trajectories from $z=3$ to $z=0$. The blue square marks the common initial position of satellite 2, its NSC, and its BH at $z=3$. Blue circles mark the positions of satellite 2 at the merger times predicted by TNG50. The yellow star and yellow curve denote the present-day position and orbital trajectory of the NSC, respectively. The black cross marks the present-day position of the central BH, and the black curve shows its trajectory since $z=3$.}
    \label{fig5}
\end{figure}

Figure~\ref{fig5} illustrates representative orbital evolutions between $z=3$ and $z=0$ for the \emph{objects} originating from satellite 2 (see blue curve in Figure~\ref{fig2}). Because GCs are initialized at different positions within the satellite, they become unbound from their host at different times, resulting in distinct orbital trajectories after their accretion into the MW (blue curves ending with blue stars). Although the BH and NSC are both initially located in the central region of the satellite, their different masses and spatial extents lead to markedly different orbital evolutions. The NSC is stripped shortly after the satellite's first pericentric passage around the MW, whereas the BH remains bound to the satellite until its complete disruption. Overall, the remnants originating from the four satellites separate into two distinct orbital families. The BHs and NSCs stripped from Sat1 and Sat2 reach large apocentres ($r_{\rm apo}\sim40\!-\!200\,\mathrm{kpc}$), whereas those originating from Sat3 and Sat4 remain confined to the inner halo ($r_{\rm apo}\sim4\!-\!10\,\mathrm{kpc}$). None of these compact remnants undergoes significant orbital decay towards the Galactic centre after being accreted. Instead, they survive as wandering BHs and NSCs orbiting within the MW halo.

\subsection{Tidal stream modelling}\label{sec:streams}

\texttt{CosmoDyn} also models the formation of tidal streams associated with compact stellar systems, including GCs, NSCs and open clusters, using a test-particle approach. GC stream modelling is activated through \texttt{ENABLE\_STREAMS=True}, while NSC streams additionally require \texttt{ENABLE\_NSC\_STREAMS=True}. The in-situ and ex-situ streams generated in the predefined execution modes follow the populations selected through \texttt{RUN\_MODE}. With \texttt{RUN\_MODE="custom"}, the GC stream stages are controlled through \texttt{RUN\_STREAM\_ICS}, \texttt{RUN\_IN\_SITU\_STREAMS}, and \texttt{RUN\_EX\_SITU\_STREAMS}. The corresponding NSC stages are controlled through \texttt{RUN\_NSC\_STREAM\_ICS}, \texttt{RUN\_IN\_SITU\_NSC\_STREAMS}, and \texttt{RUN\_EX\_SITU\_NSC\_STREAMS}.

Once the orbit of a cluster has been computed, its gravitational potential is added to the host galaxy as a \textit{MovingObjectPotential}, in the same way as satellite galaxies, to construct the full galactic potential. A \cite{Plummer} distribution of massless test particles is then integrated in the combined gravitational potential of the host galaxy and the time-dependent moving cluster potential. The cluster follows the orbit previously computed by \texttt{CosmoDyn}, while its mass evolves according to the mass-loss prescription described in Section~\ref{MLP}. Consequently, the cluster potential varies throughout the integration, whereas its Plummer scale radius is kept constant. Unless otherwise stated, each cluster is represented by \texttt{N\_STREAM\_PARTICLES$=10^5$} test particles and generated through \texttt{N\_STREAM\_ITER=30} iterations. Separate Plummer realizations are constructed for GCs and NSCs using their respective masses and half-mass radii. Because these realizations do not depend on the host galaxy, they are generated once and shared by all galaxies listed in \texttt{GALAXY\_IDS}. The moving Plummer potentials required for stream modelling are generated during the preceding dynamical stage. For GCs, this behaviour is enabled internally when stream modelling is requested. For NSCs, it is controlled through \texttt{GENERATE\_NSC\_MOVING\_POTENTIALS}. Setting \texttt{GENERATE\_NSC\_MOVING\_POTENTIALS=True} stores the evolving NSC potential along its orbit, whereas setting it to \texttt{False} omits these files and prevents the corresponding stream calculation. The implementation adopted here is inspired by the \texttt{tstrippy}\footnote{\url{https://tstrippy.readthedocs.io/en/latest/}} Python package \citep{Sal23}, which has been successfully applied to model tidal streams of both GCs \citep{Sal23,Sal25} and open clusters \citep{Casamiquela22,Parul26} in the MW. Although the test-particle approach does not reproduce realistic stellar number counts or density variations along the stream, it accurately captures the morphology of tidal streams while remaining computationally inexpensive.

Figure~\ref{fig6} presents the distribution of the stellar debris associated with all GCs accreted from the four satellite galaxies, projected in the $E$-$L_{z}$ space. In this work, tidal streams are identified through visual inspection, as our objective is not to provide a quantitative stream classification but rather to demonstrate the capability of \texttt{CosmoDyn} to reproduce the diverse morphologies of tidal streams. We therefore classify each GC according to whether it exhibits a coherent tidal stream or a fully phase-mixed stellar distribution. This visual classification is reflected in the $E$-$L_{z}$ diagram. GCs with orbital energies below approximately $E\lesssim-10^{5}\,\mathrm{km^2\,s^{-2}}$ do not exhibit coherent tidal streams because their stripped stars have already phase-mixed within the Galactic tidal field. Conversely, at the highest orbital energies, tidal streams are generally still forming and remain too short or diffuse to be clearly identified. The most prominent tidal streams are therefore found at intermediate-to-high orbital energies. No tidal streams are identified among the in-situ GC population. For the ex situ population, all GCs originating from Sat3 and Sat4 exhibit fully phase-mixed stellar distributions. In contrast, the higher-energy GCs accreted from Sat1 and Sat2 retain numerous coherent tidal streams, with 14 and 26 streams identified, respectively.

\section{Conclusion}

We have presented \texttt{CosmoDyn}, an open-source Python framework for studying the dynamics of compact stellar systems and BHs in environments reconstructed from cosmological simulations. By combining time-dependent analytical potentials with explicit orbital integrations, it bridges the gap between the resolution of cosmological simulations and the scales of GCs, NSCs, and BHs. A reconstructed environment can be reused to explore different objects, initial conditions, and physical prescriptions without rerunning the original simulation. The code, documentation, and examples are publicly available at \url{https://github.com/Blackholan/CosmoDyn}. We encourage users to report issues and feature requests, and contribute new physical modules to the continued development of \texttt{CosmoDyn}.

A major priority is to complement the current decomposition into fixed analytical density profiles with basis-function expansions, such as the Hernquist--Ostriker self-consistent-field method \citep{HernquistOstriker92}. This would provide more flexible representations of flattened, triaxial, and locally structured mass distributions while retaining the computational efficiency required for large samples. Future developments will also include more flexible prescriptions for the birth properties of compact objects. In particular, incorporating physically motivated GC-formation models, such as that developed by \citet{Leaman17}, would connect the initialization of GCs more directly to the local conditions of star formation in cosmological simulations. These improvements will preserve the modular structure of \texttt{CosmoDyn}, allowing new potentials, formation models, and dynamical prescriptions to be tested independently.

\texttt{CosmoDyn} can be applied to a broad range of studies of compact-object dynamics across different galactic environments. For MW-like galaxies, its non-spherical potentials and anisotropic dynamical-friction prescription enable the dynamics of compact objects in disc environments to be investigated. Its mass-loss and tidal-stream modules can constrain the contribution of disrupted GCs to the Galactic stellar halo \citep{Belokurov06,Horta21,Gieles23}. In MW-like galaxies, tracking accreted NSCs and BHs can also provide predictions for wandering BH populations and for intermediate-mass-ratio inspirals potentially detectable by \textit{LISA} \citep{Tremmel18,Bellovary25}. The framework can further be used to explore the influence of DM on all these populations. Including moving DM subhaloes extracted from cosmological simulations would allow their perturbations of GCs, NSCs, BHs, and stellar streams to be studied within self-consistent galaxy assembly histories, extending previous controlled studies \citep{Erkal16,Bovy17,Bonaca19,Boldrini20,Vitral26}. In alternative DM scenarios, \texttt{CosmoDyn} provides a natural extension of \citet{Boldrini26}, who considered only in-situ GCs in FDM, to ex-situ GCs, NSCs, BHs, and tidal streams. Finally, the computational efficiency of \texttt{CosmoDyn} makes it particularly well suited to the large samples of GC and NSC candidates expected from future \textit{Euclid} data releases \citep{saifollahi2025_fornax,saifollahi2025_perseus}.

\section{Data Availability}

The data underlying this article is available through reasonable request to the author. 

\begin{acknowledgements} 

PB acknowledges funding from the CNES post-doctoral fellowship program. This
work was also supported by CNES, focused on the Gaia mission. PB and PDM are grateful to the "Action Thématique de Cosmologie et Galaxies (ATCG), Programme National ASTRO of the INSU (Institut National des Sciences de l'Univers) for supporting this research, in the framework of the project "Coevolution of globular clusters and dwarf galaxies, in the context of hierarchical galaxy formation: from the Milky Way to the nearby Universe" (PI: G. Pagnini). CL acknowledges funding from the European Research Council (ERC) under the European Union’s Horizon 2020 research and innovation programme (grant agreement No. 852839). CL also acknowledges funding from the Agence Nationale de la Recherche (ANR project ANR-24-CPJ1-0160-01). This work has been done using the following software, packages and python libraries: \texttt{Numpy} (\citealp{numpy}), \texttt{Scipy} (\citealp{scipy}), \texttt{Astropy} (\citealp{astropy})

\end{acknowledgements}

\bibliography{src}

\appendix

\section{Dynamical friction models}

\begin{figure}
    \centering
    \includegraphics[width=1\linewidth]{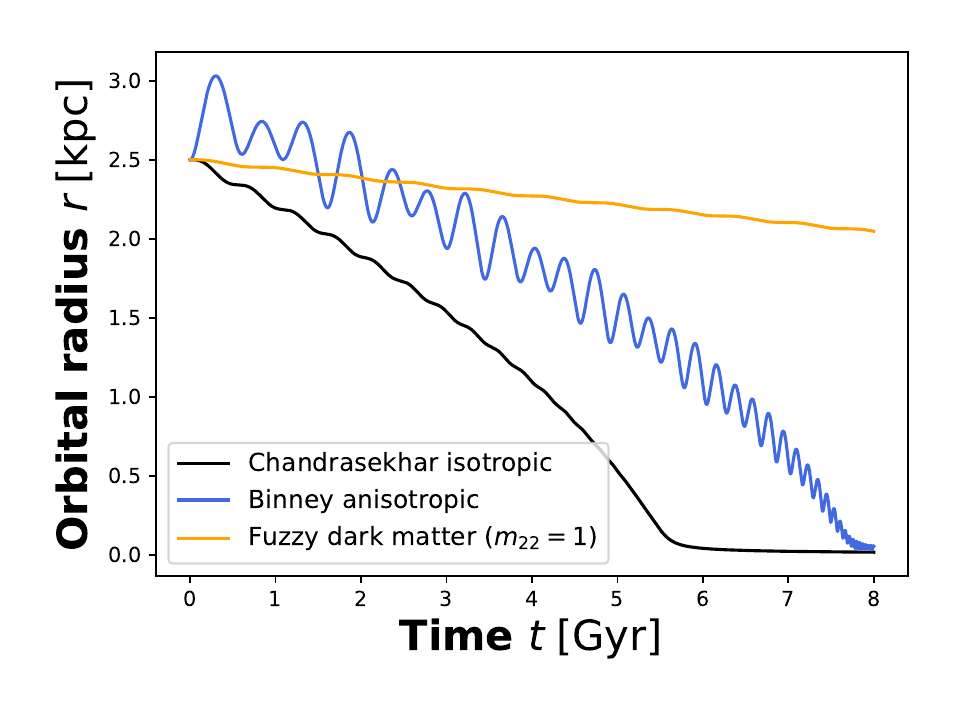}
    \caption{Orbital evolution of a $10^6\,M_\odot$ globular cluster over 8 Gyr in a $10^9\,M_\odot$ DM halo. The GC is initially placed on a circular orbit at a galactocentric distance of 2.5 kpc, with its initial velocity computed from the spherical NFW potential. The black curve shows the evolution in a spherical NFW halo including \cite{Chandra43} dynamical friction. The blue curve corresponds to an axisymmetric halo ($b=0.6$, $c=1$) including the anisotropic dynamical friction formalism of \cite{Binney77}. The orange curve shows the evolution in an FDM cored halo including FDM dynamical friction for $m_{22}=1$.}
    \label{fig4}
\end{figure}

\section{Tidal streams of globular clusters in our application}

\begin{figure*}
    \centering
    \includegraphics[width=1\linewidth]{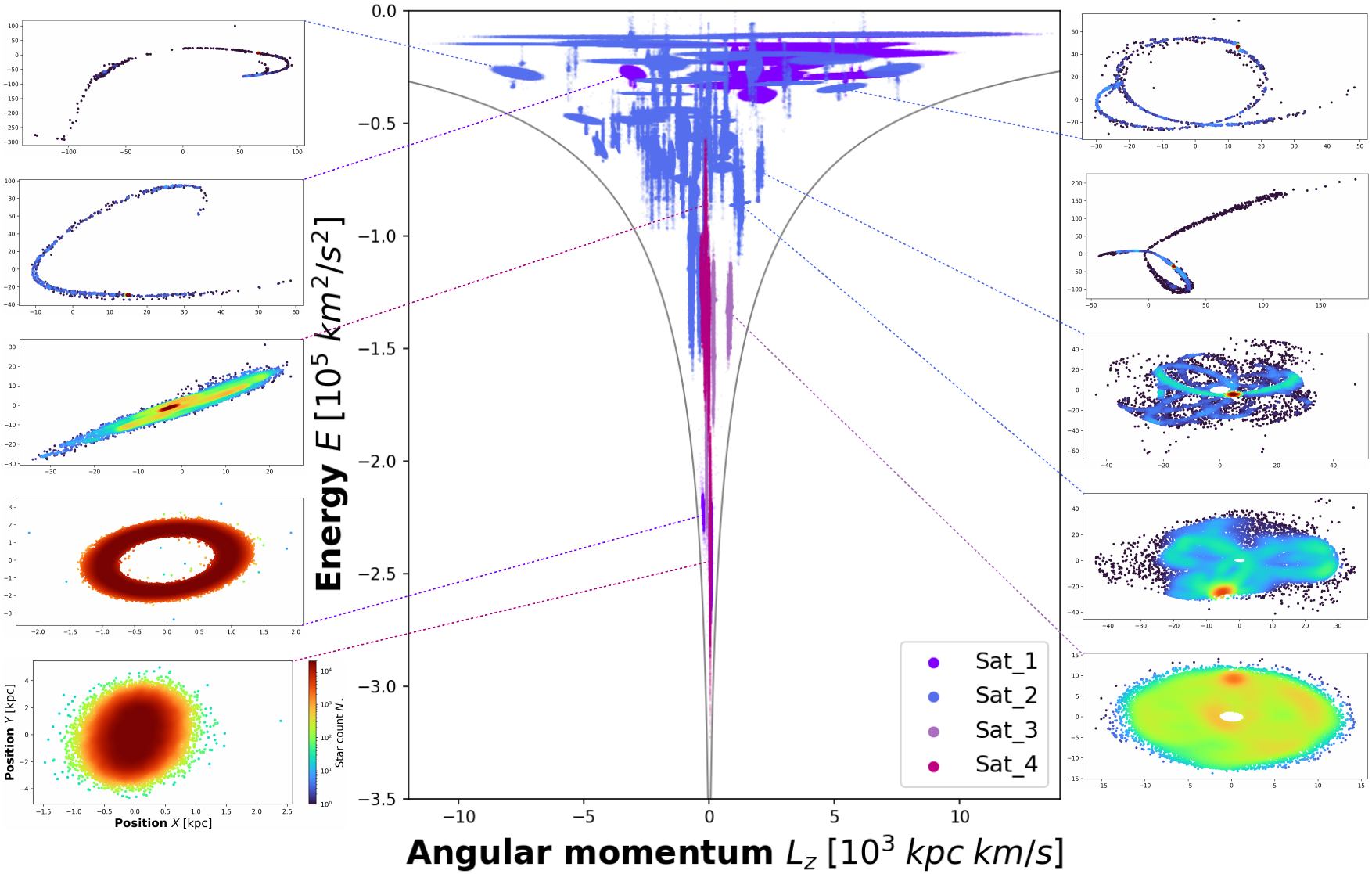}
    \caption{Distribution of stars stripped from the GCs associated with the four satellites of our best MW-like galaxy in the $L_{z}$-$E$ space at $z=0$, colour-coded by satellite. The grey curves indicate the locus of circular orbits, $L_{z,\mathrm{circ}}(E)$. The side panels show the projected spatial distribution of the GC debris associated with each satellite at $z=0$, illustrating the diversity of tidal morphologies, from thin stellar streams to fully phase-mixed structures. Dotted lines connect each debris distribution to its corresponding location in the $L_{z}$-$E$ plane.}
    \label{fig6}
\end{figure*}

\section{Inputs and outputs}\label{ApA}

Tables~\ref{tab:inputs} and \ref{tab:outputs} summarize the input files required by \texttt{CosmoDyn} and the main files generated by the pipeline, respectively. Their paths, file formats, and data structures are provided to facilitate the preparation of new runs and the interpretation of their results.

\begin{table*}
\centering
\caption{Input files required by \texttt{CosmoDyn}.}
\label{tab:inputs}
\resizebox{\textwidth}{!}{
\begin{tabular}{llll}
\hline
Input & Path & File format & Data structure \\
\hline

Construct an isolated-galaxy potential
& \texttt{examples/create\_potential.py}
& Python
& Parameters described in Appendix~X \\

Example launcher
& \texttt{examples/run\_G1\_new.py}
& Python
& Parameters described in Appendix~X \\

Complete launcher
& \texttt{examples/run\_application.py}
& Python
& Parameters described in Appendix~X \pb{Ajouter}\\

Integration times
& \texttt{TIMESTEP\_FILEG<ID>.txt}
& ASCII
& $[t_{\rm start},t_{\rm end},N_{\rm steps}]$ \\

Host-galaxy properties
& \texttt{PATH1/DataG<ID>.txt}
& ASCII
& $[
\textcolor{red}{\mathrm{Snap}_{\rm sim}},
t_{\rm sim},
\textcolor{red}{z_{\rm sim}},
\textcolor{red}{\mathrm{IdG}_{\rm sim}},
\textcolor{red}{M_{\rm dm}},
\textcolor{red}{A_{M_{\rm dm}}},
\textcolor{red}{M_\star},
\textcolor{red}{A_{M_\star}},
r_{\rm dm},
\textcolor{red}{r_{\rm s,dm}},
\textcolor{red}{r_\star},
x,y,z,v_x,v_y,v_z
]$ \\

Host-galaxy potentials
& \texttt{PATH1/PotsG<ID>.pkl}
& Pickle
& $\{\mathrm{snapshot}:
[\Phi_\star,\Phi_{\rm dm}]\}$ \\

Satellite identifiers and snapshot ranges
& \texttt{PATH1/G<ID>TimeSat.txt}
& ASCII
& \texttt{<SAT>: <snapshot 1>,...,<snapshot N>} \\

Satellite properties
& \texttt{PATH2/G<ID>Sat<SAT>.txt}
& ASCII
& $[
\textcolor{red}{\mathrm{Snap}_{\rm sim}},
t_{\rm sim},
\textcolor{red}{\mathrm{IdG}_{\rm sim}},
\textcolor{red}{M_{\rm dm}},
\textcolor{red}{A_{M_{\rm dm}}},
\textcolor{red}{M_\star},
\textcolor{red}{A_{M_\star}},
r_{\rm dm},
\textcolor{red}{r_{\rm s,dm}},
\textcolor{red}{r_\star},
x,y,z,v_x,v_y,v_z
]$\\

Satellite potentials
& \texttt{PATH2/PotsGSat<ID>N<SAT>.pkl}
& Pickle
& $\{\mathrm{snapshot}:
[\Phi_\star,\Phi_{\rm dm}]\}$ \\
\hline
\end{tabular}
}
\tablefoot{\texttt{PATH1=DataG<ID>} denotes the directory containing the input data of the host galaxy, while 
\texttt{PATH2=PATH1/GSat<ID>} denotes the directory containing the 
input data of its satellites. Here, \texttt{<ID>} and 
\texttt{<SAT>} are the host-galaxy and satellite identifiers, 
respectively. For the galaxy-property files, \texttt{Snap$_{\rm sim}$}, 
$t_{\rm sim}$, $z_{\rm sim}$, and \texttt{IdG$_{\rm sim}$} denote the simulation snapshot, cosmic time, redshift, and galaxy identifier at the corresponding epoch in the simulation, respectively. $M_{\rm dm}$ and $M_\star$ are the DM and stellar masses, while $A_{M_{\rm dm}}$ and $A_{M_\star}$ are the amplitudes of their 
corresponding analytical \texttt{galpy} potentials. The quantities 
$r_{\rm hm}^{\rm dm}$, $r_{\rm s}^{\rm DM}$, and $r_{\rm hm}^\star$ denote the half-mass DM radius, the NFW DM scale radius, and the half mass stellar radius, 
respectively. Finally, $(x,y,z)$ and $(v_x,v_y,v_z)$ are the position 
and velocity of the galaxy centre. Each row of an ASCII table corresponds to one simulation snapshot. Throughout all input files, times are expressed in Gyr, masses in 
$\mathrm{M_\odot}$, distances in kpc, and velocities in 
$\mathrm{km\,s^{-1}}$. The quantities highlighted in red represent the minimum information that must be provided in this galaxy-property file to run \texttt{CosmoDyn}. The remaining entries may be set to \texttt{NaN} (see \texttt{create\_potential.py}).}
\end{table*}

\begin{table*}
\centering
\caption{Main output files generated by \texttt{CosmoDyn}.}
\label{tab:outputs}
\resizebox{\textwidth}{!}{
\begin{tabular}{llll}
\hline
Output & Path & File format & Data structure \\
\hline

\multicolumn{4}{l}{
\textbf{Satellite trajectories and full potential 
(Section~\ref{sec:full_potential})}
}
\\
\hline

Satellite trajectories
&
\texttt{PATH3/ExSitu/Satellites/ExSituSatelliteDataG<ID>.h5}
&
HDF5
&
$t$, snapshot, positions, velocities, $M_{\rm dm}$, and $M_\star$
\\

Full host and satellite potential
&
\texttt{PATH3/ExSitu/Satellites/Potentials/FullHostPotentialG<ID>.pkl}
&
Pickle
&
$\{\mathrm{snapshot}:[\Phi_{\rm host},\Phi_{\rm sat,1},\ldots]\}$
\\

\hline
\multicolumn{4}{l}{
\textbf{Initial conditions (Section~\ref{sec:initial_conditions})}
}
\\
\hline

In-situ GC initial conditions
&
\texttt{PATH3/InSitu/IniGCG<ID>.txt}
&
ASCII
&
$[R,v_R,v_T,z,v_z,\phi]$
\\

Ex-situ GC initial conditions
&
\texttt{PATH3/ExSitu/IniG<ID>Sat<SAT>GCs.txt}
&
ASCII
&
$[R,v_R,v_T,z,v_z,\phi]$
\\

In-situ NSC or BH initial conditions
&
\texttt{PATH3/<OBJECT>/InSitu/Ini<OBJECT>G<ID>.txt}
&
ASCII
&
$[R,v_R,v_T,z,v_z,\phi]$; BH files also contain $M_{\rm BH}$
\\

Ex-situ NSC or BH initial conditions
&
\texttt{PATH3/<OBJECT>/ExSitu/IniG<ID>Sat<SAT><OBJECT>.txt}
&
ASCII
&
$[R,v_R,v_T,z,v_z,\phi]$; BH files also contain $M_{\rm BH}$
\\

\hline
\multicolumn{4}{l}{
\textbf{Orbital dynamics (Section~\ref{sec:dynamics})}
}
\\
\hline

In-situ GC dynamics
&
\texttt{PATH3/InSitu/InSituDynamics\_<CONFIG>\_G<ID>.h5}
&
HDF5
&
$t$, phase-space coordinates, mass, snapshot, and capture status
\\

Ex-situ GC dynamics
&
\texttt{PATH3/ExSitu/ExSituDynamics\_<CONFIG>\_G<ID>.h5}
&
HDF5
&
$t$, phase-space coordinates, mass, bound status, and release information
\\

In-situ NSC or BH dynamics
&
\texttt{PATH3/<OBJECT>/InSitu/InSitu<OBJECT>Dynamics\_<CONFIG>\_G<ID>.h5}
&
HDF5
&
$t$, phase-space coordinates, mass, snapshot, and capture status
\\

Ex-situ NSC or BH dynamics
&
\texttt{PATH3/<OBJECT>/ExSitu/ExSitu<OBJECT>Dynamics\_<CONFIG>\_G<ID>.h5}
&
HDF5
&
$t$, phase-space coordinates, mass, bound status, and release information
\\

\hline
\multicolumn{4}{l}{
\textbf{Tidal streams (Section~\ref{sec:streams})}
}
\\
\hline

In-situ GC or NSC streams
&
\texttt{PATH3/.../Streams\_<CONFIG>/InSituStream\_<OBJECT><N>.h5}
&
HDF5
&
Final positions, velocities, and bound status of the stream particles
\\

Ex-situ GC or NSC streams
&
\texttt{PATH3/.../Streams\_<CONFIG>/ExSituStream\_G<ID>\_Sat<SAT>\_<OBJECT><N>.h5}
&
HDF5
&
Final positions, velocities, and release status of the stream particles
\\

\hline
\multicolumn{4}{l}{
\textbf{Diagnostics and computational performance}
}
\\
\hline

Diagnostic figures
&
Corresponding \texttt{Plots/} directory
&
PNG
&
Orbital-radius evolution or projected stream distribution
\\

Computation-time report
&
\texttt{Outputs/ComputationTime\_<CONFIG>.txt}
&
ASCII
&
Stage, galaxy identifier, execution time, object number, and CPU number
\\

\hline
\end{tabular}
}
\tablefoot{\texttt{PATH3=Outputs/G<ID>} denotes the output directory of the host galaxy. Here, \texttt{<ID>}, \texttt{<SAT>}, and \texttt{<OBJECT>} denote the host-galaxy identifier, satellite identifier, and object type, respectively. The tag \texttt{<CONFIG>} identifies the potential, moving-satellite, dynamical-friction, and mass-loss configuration. Throughout all input and output files, times are expressed in Gyr, masses in $\mathrm{M_\odot}$, distances in kpc, velocities in 
$\mathrm{km\,s^{-1}}$, and angles in radians. Phase-space coordinates 
are given in the cylindrical \texttt{galpy} convention 
$(R,v_R,v_T,z,v_z,\phi)$.}
\end{table*}

\section{Getting started}\label{ApB}
\label{app:getting_started}

This appendix provides three complementary entry points to \texttt{CosmoDyn}. New users are encouraged to begin with the minimal example based on \texttt{create\_potential.py}, \texttt{TimeStepG1.txt}, and \texttt{run\_G1\_new.py}, which shows how to construct the potential of an isolated galaxy and integrate a small population of in-situ GCs. Users wishing to test the complete pipeline rapidly, including ex-situ objects and tidal streams, should then use \texttt{run\_application\_low\_resolution.py}. Finally, \texttt{run\_application.py} contains the full configuration adopted for the TNG50 application presented in this work and should be used together with the supplied input files for galaxy ID 613192 to reproduce our calculations. All scripts and associated files are available in the public repository at \url{https://github.com/Blackholan/CosmoDyn}.

\subsection{Installation}

To avoid conflicts with existing Python packages, we recommend installing \texttt{CosmoDyn} in a dedicated virtual environment. The repository can then be downloaded and installed in editable mode as follows:
\begin{lstlisting}[style=pythonstyle]
git clone https://github.com/Blackholan/CosmoDyn.git
cd CosmoDyn

python3 -m venv cosmodyn_env
source cosmodyn_env/bin/activate

python3 -m pip install -e .
\end{lstlisting}

\subsection{Constructing an isolated-galaxy potential}

The file \texttt{create\_potential.py} provides a simple example for 
constructing an analytical galactic potential without using a 
cosmological simulation. The user first specifies the galaxy properties 
at the beginning of the file:
\begin{lstlisting}[style=pythonstyle]
GALAXY_ID = 1
DM_HALO_MASS = 1e9       # Msun
STELLAR_MASS = 1e7       # Msun
REDSHIFT = 0.0
STELLAR_SCALE_RADIUS = 0.8  # kpc
\end{lstlisting}
In this example, the DM halo is represented by an NFW potential. 
Its concentration and scale radius are computed with 
\texttt{COLOSSUS} \citep{Colossus}, while the stellar component is represented by a 
Hernquist potential. The user may retain only the DM component 
or return both components as a list:
\begin{lstlisting}[style=pythonstyle]
return dark_matter_potential
\end{lstlisting}
or
\begin{lstlisting}[style=pythonstyle]
return [stellar_potential, dark_matter_potential]
\end{lstlisting}
The example potential is generated by running
\begin{lstlisting}[style=pythonstyle]
python3 examples/create_potential.py
\end{lstlisting}
For \texttt{GALAXY\_ID=1}, this command creates the directory 
\texttt{DataG1/} containing the two files required by 
\texttt{CosmoDyn}:
\begin{itemize}
    \item \texttt{DataG1/DataG1.txt}, containing the structural 
    properties of the galaxy;
    \item \texttt{DataG1/PotsG1.pkl}, containing the analytical 
    \texttt{galpy} potential.
\end{itemize}
The exact structure and units of these files are given in 
Table~\ref{tab:inputs}. More complex or time-dependent potentials can 
be introduced by storing one set of galaxy properties and one 
\texttt{galpy} potential for each snapshot.

\subsection{Defining the integration times}

The temporal sampling is provided through \texttt{TimeStepG1.txt}. Each 
row follows the format
\begin{lstlisting}[style=pythonstyle]
t_start  t_end  N_steps
\end{lstlisting}
where the two times are expressed in Gyr and 
\texttt{N\_steps} is the number of samples used over the corresponding 
interval. The example file follows the isolated galaxy over 
$13.803\,\mathrm{Gyr}$:
\begin{lstlisting}[style=pythonstyle]
0.0  13.803  2000
\end{lstlisting}

\subsection{Generating and integrating in-situ GCs}

The minimal calculation is launched with \texttt{run\_G1\_new.py}. The 
galaxy identifier and input timestep file must match those created in 
the previous steps:
\begin{lstlisting}[style=pythonstyle]
GALAXY_IDS = [1]
RUN_MODE = "in_situ"

SNAPSHOT_INDEX = 0
END_SNAPSHOT_INDEX = None

TIMESTEP_FILE = "TimeStepG1.txt"
POTENTIAL_MODE = "static"
STATIC_POTENTIAL_INDEX = 0
\end{lstlisting}
The GC population is then defined through, for example,
\begin{lstlisting}[style=pythonstyle]
NGC = 3
CIRCULARITY_THRESHOLD = None
TAGGING_RADIUS_FACTOR = 2
MINIMUM_TAGGING_RADIUS = 0.4

GC_MASS = 1e6
GC_HALF_MASS_RADIUS = 0.01
\end{lstlisting}
Here, \texttt{NGC=3} directly generates three GCs. Alternatively, 
setting \texttt{NGC=0} computes their number from the halo-mass 
relation described by Equation~\ref{eq:ngc_relation}. Setting 
\texttt{CIRCULARITY\_THRESHOLD=None} disables the orbital-circularity 
selection. For a first orbital calculation without tidal streams, we recommend
\begin{lstlisting}[style=pythonstyle]
ENABLE_STREAMS = False
\end{lstlisting}
The complete example is then executed with
\begin{lstlisting}[style=pythonstyle]
python3 examples/run_G1_new.py
\end{lstlisting}
The generated GC initial conditions and dynamical evolution are saved in
\begin{lstlisting}[style=pythonstyle]
Outputs/G1/InSitu/IniGCG1.txt
Outputs/G1/InSitu/InSituDynamics_<CONFIG>_G1.h5
\end{lstlisting}
and the corresponding orbital-radius diagnostic is saved in the 
\texttt{Outputs/G1/InSitu/Plots/} directory. Tidal streams can 
subsequently be included by setting \texttt{ENABLE\_STREAMS=True}. 
This minimal workflow can be generalized to a time-dependent 
cosmological environment by replacing the isolated-galaxy inputs with 
the host and satellite files described in Table~\ref{tab:inputs}.

\subsection{Launcher used for the application presented in this work}\label{}

The complete launcher used to produce the application presented in this work, named \texttt{run\_application.py}, is reproduced below and is also available in the public \texttt{CosmoDyn} repository at \url{https://github.com/Blackholan/CosmoDyn/tree/main/examples}. Together with the input files described in Table~\ref{tab:inputs}, it provides all the physical and numerical parameters required to reproduce our calculation. We also provide all the input files required to reproduce the analysis of our TNG50 MW-like galaxy (ID=613192) at \url{https://github.com/Blackholan/CosmoDyn/tree/main/DataG613192}. The comments in the launcher document the alternative options available in \texttt{CosmoDyn}, including those not activated in this application. Nevertheless, we encourage users to explore the code and its full range of functionalities using the \texttt{run\_application\_low\_resolution.py} script. This computationally inexpensive configuration uses a reduced number of objects (\texttt{NGC=2}) and stream particles (\texttt{N\_STREAM\_PARTICLES=100}), allowing the complete pipeline to be tested rapidly.

\begin{lstlisting}[
    style=pythonstyle,
    caption={Complete \texttt{CosmoDyn} launcher used for the application 
    presented in this work.},
    label={lst:complete_launcher}
]
#!/usr/bin/env python3
# coding: utf-8
import numpy as np
from cosmodyn.pipeline import run_pipeline

# ==========================================================
# 1. RUN SELECTION
# ==========================================================

GALAXY_IDS = [613192] # [613192, 462710]
# GALAXY_IDS = np.loadtxt("GalaxyList.txt", dtype=int).tolist()
CONTINUE_ON_ERROR = True # Continue with the remaining galaxies if one galaxy fails
RUN_MODE = "custom" # "in_situ", "ex_situ", "full", "custom"
ENABLE_STREAMS = True
ENABLE_NSC_STREAMS = False

# ==========================================================
# 2. INITIAL CONDITIONS
# ==========================================================

SNAPSHOT_INDEX = 0
NGC = 0                  # 0: the halo-mass relation; >0: the number directly
ALPHA = 3                # used only when NGC = 0
CIRCULARITY_THRESHOLD = 0.6  # Minimum Lz/Lcirc; None disables the circularity selection
TAGGING_RADIUS_FACTOR = 1 # Maximum tagging radius in units of the stellar half-mass radius
MINIMUM_TAGGING_RADIUS = None  # Minimum tagging radius in kpc; None disables this lower limit

# AGAMA parameters
N_ITER = 20
N_PARTICLES_PER_COMPONENT = 100_000
RANDOM_SEED = None    
KEEP_AGAMA_FILE = True # Keep the generated AGAMA particle file for reuse

# ==========================================================
# NUCLEAR STAR CLUSTERS
# ==========================================================

ENABLE_NSCS = True

NSC_INITIAL_RADIUS = 0.001          # kpc 
NSC_MASS = 0                      # Msun
NSC_HALF_MASS_RADIUS = 0.01        # kpc 
NSC_CENTRAL_CAPTURE_RADIUS = 0.0001 # kpc
GENERATE_NSC_MOVING_POTENTIALS = False

# ==========================================================
# BLACK HOLES
# ==========================================================
ENABLE_BHS = True

BH_MASS = 0                         # Msun
BH_INITIAL_RADIUS = 0.001           # kpc 
BH_CENTRAL_CAPTURE_RADIUS = 0.0001  # kpc

# ==========================================================
# 3. DYNAMICS FOR IN-SITU AND EX-SITU GCs
# ==========================================================

END_SNAPSHOT_INDEX = None # Final snapshot index; None integrates to the last available snapshot
INTEGRATION_METHOD = "dop853_c"
TIMESTEP_FILE = "TimeStepGTNG50.txt"

POTENTIAL_MODE = "evolving"  # "evolving" or "static"
STATIC_POTENTIAL_INDEX = 73 # used only with POTENTIAL_MODE = "static"
INCLUDE_MOVING_SATELLITES = True # include the gravitational field of moving satellite galaxies

DF_MODEL = "cdm"             # "none", "cdm", "binney", "fdm"
M22 = 1                       # used only with DF_MODEL = "fdm"

GC_MASS = 1e6                  # Msun
GC_HALF_MASS_RADIUS = 0.01     # kpc
CENTRAL_CAPTURE_RADIUS = 0.001  # kpc
MASS_LOSS_MODE = "coupled" # "none", "postprocess" or "coupled"

# ==========================================================
# 4. OPTIONAL EX-SITU OVERRIDES
# Ignored in RUN_MODE = "in_situ".
# ==========================================================

NGC_EX_SITU = NGC
ALPHA_EX_SITU = ALPHA
EX_SITU_CIRCULARITY_THRESHOLD = CIRCULARITY_THRESHOLD
EX_SITU_TAGGING_RADIUS_FACTOR = TAGGING_RADIUS_FACTOR
EX_SITU_MINIMUM_TAGGING_RADIUS = MINIMUM_TAGGING_RADIUS
MAXIMUM_SATELLITE_RADIUS = 1000.0  # kpc
RELEASE_ENERGY_TOLERANCE = 0.0 # km^2 s^-2


# ==========================================================
# 5. OPTIONAL STREAM PARAMETERS
# Used only when ENABLE_STREAMS = True.
# ==========================================================

N_STREAM_PARTICLES = 100000
N_STREAM_ITER = 30 # AGAMA parameter to build the Plummer model
OVERWRITE_STREAM_ICS = False # True regenerates existing Plummer initial conditions
STREAM_N_JOBS = -1
STREAM_BATCH_SIZE = 64

# ==========================================================
# 6. CUSTOM STAGES USED ONLY WITH RUN_MODE = "custom"
# ==========================================================

#Globular clusters
RUN_ICS = True
RUN_EX_SITU_ICS = True
RUN_EX_SITU_SATELLITES = True

RUN_IN_SITU_DYNAMICS = True
RUN_EX_SITU_DYNAMICS = True

RUN_STREAM_ICS = True
RUN_IN_SITU_STREAMS = True
RUN_EX_SITU_STREAMS = True

#Nuclear star clusters
RUN_IN_SITU_NSC_ICS = False
RUN_EX_SITU_NSC_ICS = True

RUN_IN_SITU_NSC_DYNAMICS = False
RUN_EX_SITU_NSC_DYNAMICS = True

RUN_NSC_STREAM_ICS = False
RUN_IN_SITU_NSC_STREAMS = False
RUN_EX_SITU_NSC_STREAMS = False

#Black holes
RUN_IN_SITU_BH_ICS = False
RUN_EX_SITU_BH_ICS = True
RUN_IN_SITU_BH_DYNAMICS = False
RUN_EX_SITU_BH_DYNAMICS = True


if __name__ == "__main__":
    run_pipeline(locals())
\end{lstlisting}

\end{document}